\documentclass[showpacs,twocolumn]{revtex4}

\usepackage{graphics}
\usepackage{graphicx}

\def\m@thcombine#1#2{%
  \setbox0=\hbox{$#1$}
  \setbox1=\hbox{$#2$}
  \ifdim\wd0>\wd1
    \setbox0=\hbox to\wd1{\hss\box0\hss}
  \else
    \setbox1=\hbox to\wd0{\hss\box1\hss}
  \fi
  \mathop{\vcenter{
    \offinterlineskip\box0\box1}}}
\def\lesim{\m@thcombine<\sim}
\def\gesim{\m@thcombine>\sim}

\newcommand{\vecr}{\mbox{\boldmath $r$}}

\newcommand{\rhot}{\tilde{\rho}}

\begin{document}

\title
{Anomalous pairing vibration
in neutron-rich Sn isotopes beyond the $N=82$ magic number
}

\author{Hirotaka Shimoyama, Masayuki Matsuo}

\affiliation{
Department of Physics, Faculty of Science and 
Graduate School of Science and Technology, 
Niigata University, Niigata 950-2181, Japan 
}

\date{\today}

\begin{abstract}
Two-neutron transfer  associated with the pair correlation
in superfluid neutron-rich nuclei is studied with focus on
 low-lying $0^+$ states in Sn isotopes beyond the $N=82$ magic number.
We describe microscopically  the two-neutron addition and
removal transitions by means of the  
Skyrme-Hartree-Fock-Bogoliubov mean-field model and the continuum quasiparticle
random phase approximation formulated in the coordinate space representation.
It is found that the
pair transfer strength for the transitions between the ground states becomes 
significantly large for the isotopes with $A \ge 140$, reflecting very
small neutron separation energy and long tails of the weakly bound
$3p$ orbits. 
In $^{132-140}$Sn,  a peculiar feature of the pair transfer is seen in 
transitions to low-lying excited $0^+$ states. They can be regarded as
a novel kind of pair vibrational  
mode which is characterized  by 
an anomalously
long tail of the
transition density extending to far outside of the
nuclear surface, and a large strength comparable to that of the
ground-state transitions.  The presence of the weakly bound neutron orbits
plays a central role for these anomalous behaviors.
\end{abstract}

\pacs{21.10.Pc, 21.10.Re, 21.60.Jz, 25.40.Hs, 27.60.+j}%

\maketitle

\section{Introduction}\label{Intro}

Recently two-neutron transfer processes  have
attracted renewed interests thanks to the 
 opportunities of two-neutron transfer reaction experiments using the
beams of radioactive neutron-rich isotopes\cite{Keeley07,Tanihata08,Chatterjee08,Golovkov09,Lemasson11}. 
In the past
a considerable number of experimental
and theoretical studies of two-neutron transfer
have been accumulated  for stable nuclei\cite{Yoshida62,Bes-Broglia66,Broglia73,BM2,Igarashi91,Oertzen-Vitturi,Brink-Broglia}. 
One of the central concepts established there
 is that the two-neutron transfer amplitude is influenced by  collective
excitation modes caused by the 
superfluidity or the pair correlation\cite{Bes-Broglia66,BM2,Broglia73,Brink-Broglia}. 
In the case of the open-shell nuclei, the strong two-neutron transfer 
transitions connecting the ground states emerge; it is a Nambu-Goldstone mode
related to the rotational symmetry with respect to the phase of the pair condensate,
called the pairing rotation. The pairing vibration, which corresponds to a vibration of the pairing gap,
is another collective mode producing a low-lying excited $0^+$ state, but the intensity of 
the associated pair transfer is significantly smaller than the pairing rotation.

   When neutron-rich nuclei are concerned, the above conventional
picture of the pairing collectivity may be modified. We can expect this because neutron-rich nuclei
often accompany skin or halo, i.e., low density distributions of weakly bound neutrons surrounding the
nucleus, and also because
the neutron pairing in neutron matter is predicted to be strong or close to the strong-coupling
regime at low densities
\cite{Esbensen,Lombardo-Schulze,Dean03,Matsuo06,Margueron08,AbeSeki09,GezerlisCarlson}. 
The expected enhancement of the pair correlation 
may result in unusual properties in two-neutron transfers.  
Dobaczewski et al.\cite{DobHFB2} 
has pointed out the possibility that the two-neutron transfer associated
with the pairing rotation may be enhanced in neutron-rich nuclei because
of the surface enhanced pairing.  
Influences of the surface enhancement on the pair vibrational 
two-neutron transfer 
populating  excited $0^+$ states and the first $2^+$ states in neutron-rich Sn isotopes 
are discussed in
Refs.~\cite{Khan09,Pllumbi10}
and in Ref.~\cite{Matsuo10}, respectively.    
Recent experiments of $(p,t)$ reaction on light-mass neutron-rich
nuclei such as  $^{11}$Li \cite{Tanihata08,Potel1,Potel2} and $^{6,8}$He 
\cite{Keeley07,Chatterjee08,Golovkov09,Lemasson11} 
point to the importance of the pair correlation
in these typical halo or skin nuclei. The giant pairing vibration in neutron-rich
nuclei has also been discussed\cite{Khan04,Avez}.

In the present paper we study the pairing collective modes in heavy-mass neutron-rich nuclei
in order to explore new features in the pairing rotation, the
pairing vibration,
and the associated two-neutron transfer amplitudes. 
We focus on  excitation modes with monopole multipolarity ($L=0$), for which 
the pairing collectivity is expected most strongly. 
As a target of the analysis, we choose the Sn isotopic chain ranging from the proton-rich
side $^{100}$Sn ($N=50$) to  very neutron-rich isotopes with $A\sim 150$ beyond  
the  $N=82$ magic number ($^{132}$Sn ). Our theoretical tool is 
the Hartree-Fock-Bogoliubov mean-field model using the Skyrme energy density
functional  and the continuum quasiparticle random phase approximation, 
and it is the same 
as adopted in Ref.\cite{Matsuo10}. 
A similar approach is made in Khan et al.\cite{Khan04,Khan09},
where the $L=0$ pairing vibration in neutron-rich
Sn isotopes is also studied. Compared with Ref.\cite{Khan09},
 we perform more systematic and detailed analyses,
which eventually bring us a new finding and some differences. Among all,
the present analysis predicts  a novel type of the pairing vibration that emerges 
in $^{134-140}$Sn.

\section{Skyrme-Hartree-Fock-Bogoliubov mean-field plus QRPA approach}

\subsection{The model}\label{sec:model}

We describe the neutron pair correlation by means of the 
Hartree-Fock-Bogoliubov (HFB) mean-field theory\cite{Bender} and the continuum
quasiparticle random phase approximation (QRPA)\cite{Matsuo01,Serizawa09}. We assume the spherical symmetry
of the mean-fields and densities associated with the ground states of
the subshell-closed Sn isotopes. The model adopted in the present study
is the same as that  in Ref.\cite{Matsuo10}.

The starting point of the model is the energy density functional, which 
we construct from the Skyrme interaction with the parameter set SLy4\cite{CBH98},
 and the 
density-dependent delta  interaction (DDDI) adopted as an 
effective pairing force\cite{Esbensen,Matsuo06,DobHFB2,
DDpair-Chas,Garrido,Garrido01,
DD-Dob,DD-mix1,DD-mix2,Hagino05,Matsuo07,Yam05,Yam09}.
The DDDI is given by
\begin{equation}\label{eqn:DDDI}
v^{{\rm pair}}_q(\vecr,\vecr')={1\over2}(1-P_\sigma)V_q(\vecr)
\delta(\vecr-\vecr') \ \ \ (q=n,p)
\end{equation}
where $V_q(\vecr)$ is the pairing interaction strength and is a function of
the neutron and proton densities.  We adopt the following form
\begin{equation}
V_n(\vecr)=v_0 \left(1-\eta\left(\frac{\rho_n(\vecr)}{\rho_c}\right)^\alpha\right)
\end{equation}
with the parameters
$v_0=-458.4$ MeV fm$^{3}$, $\rho_c=0.08$fm$^{-3}$, $\alpha=0.59$, and 
$\eta=0.71$ \cite{Matsuo06,Matsuo07,Matsuo10}.
Here the parameter $v_0$ representing the strength in the
free space is 
chosen to reproduce the scattering length $a=-18.5$ fm 
of the bare $nn$-interaction in the $^{1}S$ channel, and 
$\alpha$ is to fit the density-dependent BCS pairing gap 
of neutron matter. The parameter $\eta$ is adjusted to
reproduce the experimental pair gap in $^{120}$Sn. We call this parameter
set DDDI-bare' as it is motivated by the bare nuclear force.
This interaction has a strong density dependence.
For comparison, we also use the volume pairing interaction
($v_0=-195$ MeV fm$^{3}$ and $\eta=0$) with no density-dependence,
and also the mix pairing interaction\cite{DD-mix1,DD-mix2}  ( $v_0=-292$ MeV fm$^{3}$)
with intermediate density dependence. See Ref.~\cite{Matsuo10}
for details. The parameter $v_0$ of the volume and the mix pairing interactions
is determined to reproduce the pair gap in $^{120}$Sn.

The HFB equation is solved  in the coordinate space
representation using the polar coordinate system. The radial
coordinate space is truncated  at $r_{{\rm max}}=20$ fm. The density
and the pair density of neutrons or protons are given as sums of the
contributions from the quasiparticle states, which we 
truncate with respect to the angular momentum
quantum numbers and the quasiparticle energy. We actually set 
 the maximum orbital angular quantum number
$l_{\rm max}=12$ and the maximum quasiparticle energy $E_{\rm max}=60$
MeV.

The excited states are described by means of the QRPA formulated 
in the coordinate space representation and on
the basis of the self-consistent HFB solution. 
The residual interactions to be used in the QRPA are derived from the
Skyrme energy density functional and the DDDI, but the
Landau-Migdal approximation is employed for the residual interaction in 
the particle-hole channel. 
We impose the outgoing-wave boundary condition on the
continuum quasiparticle states as described in Ref.\cite{Matsuo01,Serizawa09} in the case of the isotopes 
with $A\ge 132$ while the box boundary condition is adopted\cite{Matsuo10}
for $A<132$ where the neutron Fermi energy is deeper than
$-6.8$ MeV. In the QRPA response functions, the smearing parameter 
$\epsilon = 50$ keV is introduced so that the obtained strength function
is convoluted with the Lorentzian with the FWHM$=2\epsilon=100$ keV.

\subsection{Strength function and transition density for
 pair transfer modes}

In the present work, we define the strength 
of the two-neutron transfer using matrix elements of the creation and annihilation
operators  of  a $S=0$ pair of neutrons with the angular momentum $L$:
\begin{eqnarray}
P^\dagger_{LM}
&=& \int d\vecr Y_{LM}(\hat{r})f(r)
\psi^\dagger(\vecr\downarrow)\psi^\dagger(\vecr\uparrow),\\
P_{LM}
&=& \int d\vecr Y^*_{LM}(\hat{r})f(r)
\psi_q(\vecr\uparrow)\psi_q(\vecr\downarrow). 
\end{eqnarray}
When we consider  transitions from the ground state of a nucleus with even $N$ to
 states in the neighboring nucleus with $N+2$ through the addition of a neutron
pair, we evaluate 
the strength function 
$P^\dagger_{LM}$ 
\begin{equation}
S_{{\rm Pad}L}(E) \equiv  \sum_{iM} \delta(E-E_{iL}) |\langle \Psi_{iLM} |P^\dagger_{LM}|\Psi_0\rangle|^2,
\end{equation}
and for transitions via the removal of a neutron pair  we calculate the strength function
\begin{equation}
S_{{\rm Prm}L}(E)\equiv  \sum_{iM} \delta(E-E_{iL}) |\langle \Psi_{iL-M} |P_{LM}|\Psi_0\rangle|^2.
\end{equation}
Here $\Psi_{0}$ is the ground state with $N$ while $\Psi_{iLM}$ is a state 
with the angular quantum numbers $LM$ and the neutron number $N\pm 2$,  
and $E_{iL}$ is the transition energy measured from the ground state of the residue nucleus
with $N\pm 2$.

For a specific transition to a given final state populated via the pair-addition
and removal operators, 
the transition densities 
\begin{eqnarray}
P_{i}^{({\rm ad})}(\vecr)&\equiv&
\langle \Psi_{iLM} |
\psi^\dagger(\vecr\downarrow)\psi^\dagger(\vecr\uparrow)| \Psi_0\rangle  
\nonumber\\
&=&
Y_{LM}^*(\hat{\vecr})P_{iL}^{({\rm ad})}(r), \\
P_{i}^{({\rm rm})}(\vecr) &\equiv&
\langle \Psi_{iLM} | \psi(\vecr\uparrow)\psi(\vecr\downarrow)| \Psi_0\rangle 
\nonumber\\
&=&
Y_{LM}^*(\hat{\vecr})P_{iL}^{({\rm rm})}(r)
\end{eqnarray}
may be defined. 
We also evaluate the reduced transition probabilities  defined by
\begin{eqnarray}
B({\rm Pad}L;{\rm gs}\rightarrow i)&\equiv &\sum_{M} |\langle \Psi_{iLM}
 |P^\dagger_{LM}|\Psi_0\rangle|^2  \nonumber\\
&=& (2L+1) \left| \int r^2 P_{iL}^{({\rm ad})}(r) dr \right|^2, \\
B({\rm Prm}L;{\rm gs}\rightarrow i)&\equiv &\sum_{M} |\langle \Psi_{iL-M} |P_{LM}|\Psi_0\rangle|^2 
\nonumber\\
&=& (2L+1) \left| \int r^2 P_{iL}^{({\rm rm})}(r) dr \right|^2.
\end{eqnarray}
For the procedure to evaluate the strength functions, the transition densities
and the transition strengths in the QRPA scheme, we refer the readers to 
 Refs.~\cite{Matsuo01,Serizawa09,Matsuo10}.

\section{Ground state properties and pairing rotation} \label{sec:gs}

\begin{figure*}[t]
\includegraphics[scale=0.3,angle=-90]{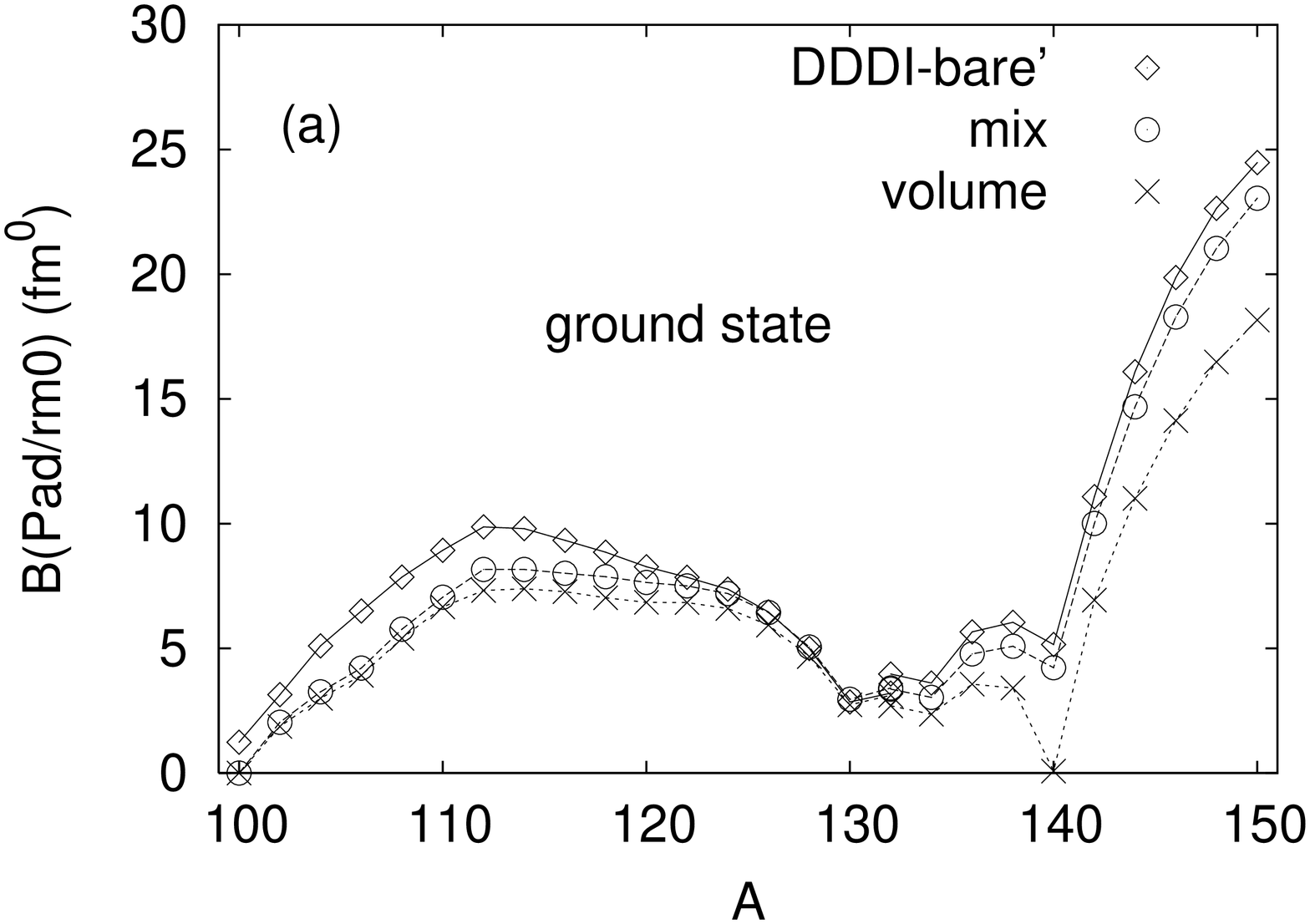}
\includegraphics[scale=0.3,angle=-90]{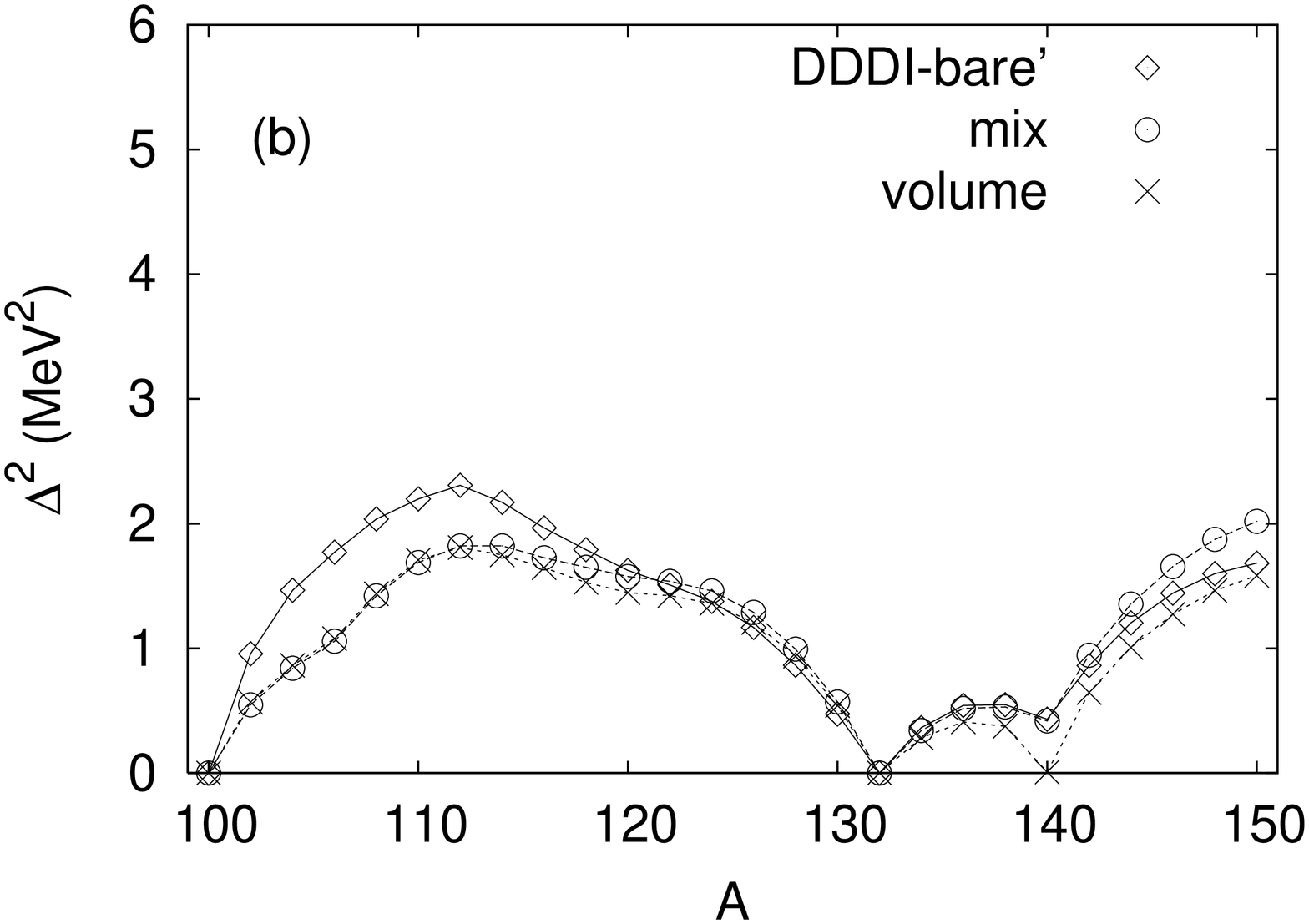}
\caption{\label{fig1}
(a) The neutron pair transfer strength $B({\rm Pad/rm}0;{\rm gs} \rightarrow {\rm gs})$ of the ground-state
transitions calculated for the even-even Sn isotopes.
 The diamonds connected with the
 solid line are the results obtained with the pairing interaction DDDI-bare' while 
the circles and the crosses are those with the mix and the
volume pairing interactions, respectively.
(b)The squared average pairing gap $\Delta_{uv}^2$ of neutrons.
} 
\end{figure*}

\begin{figure*}[t]
\includegraphics[scale=0.3,angle=-90]{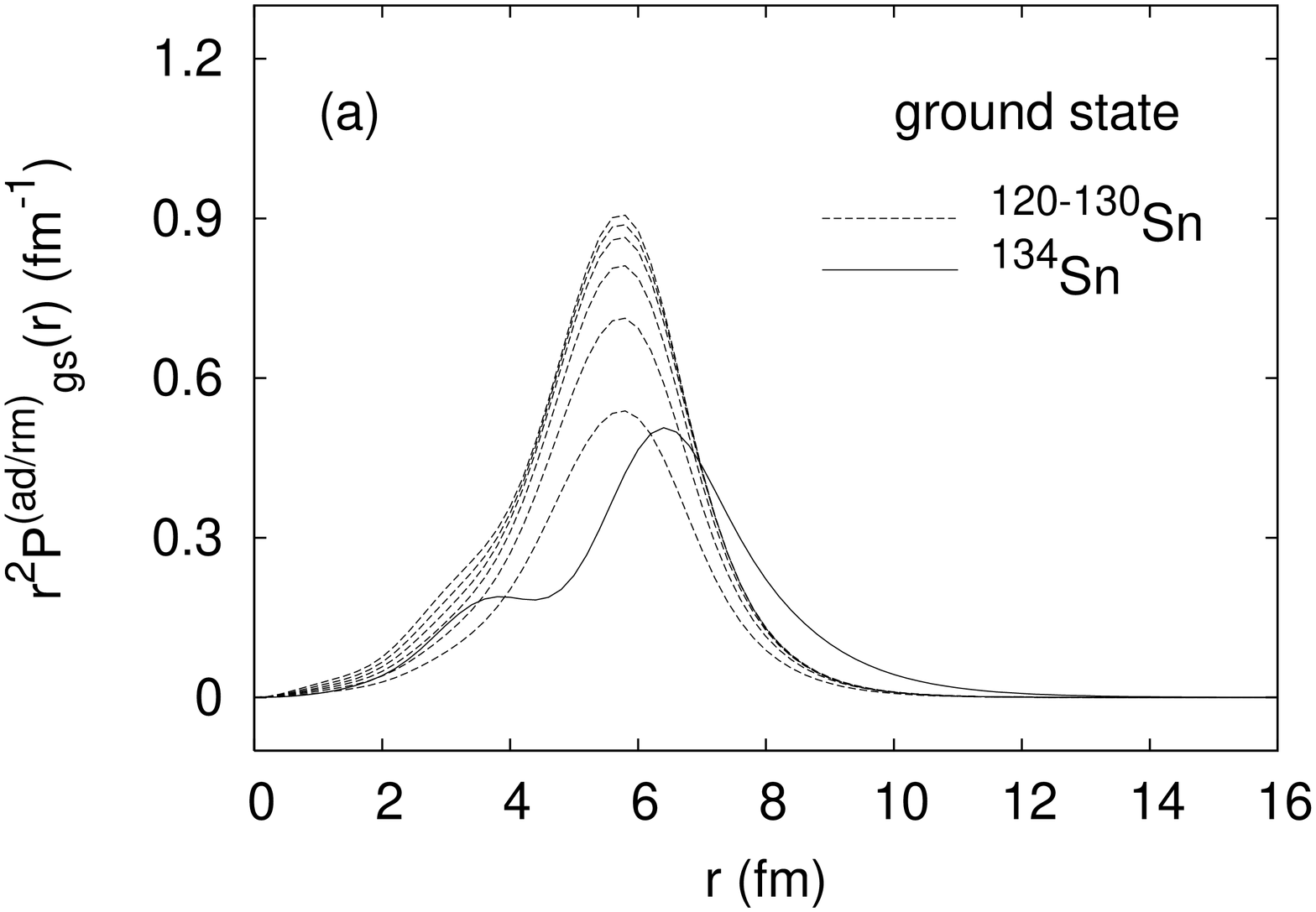}
\includegraphics[scale=0.3,angle=-90]{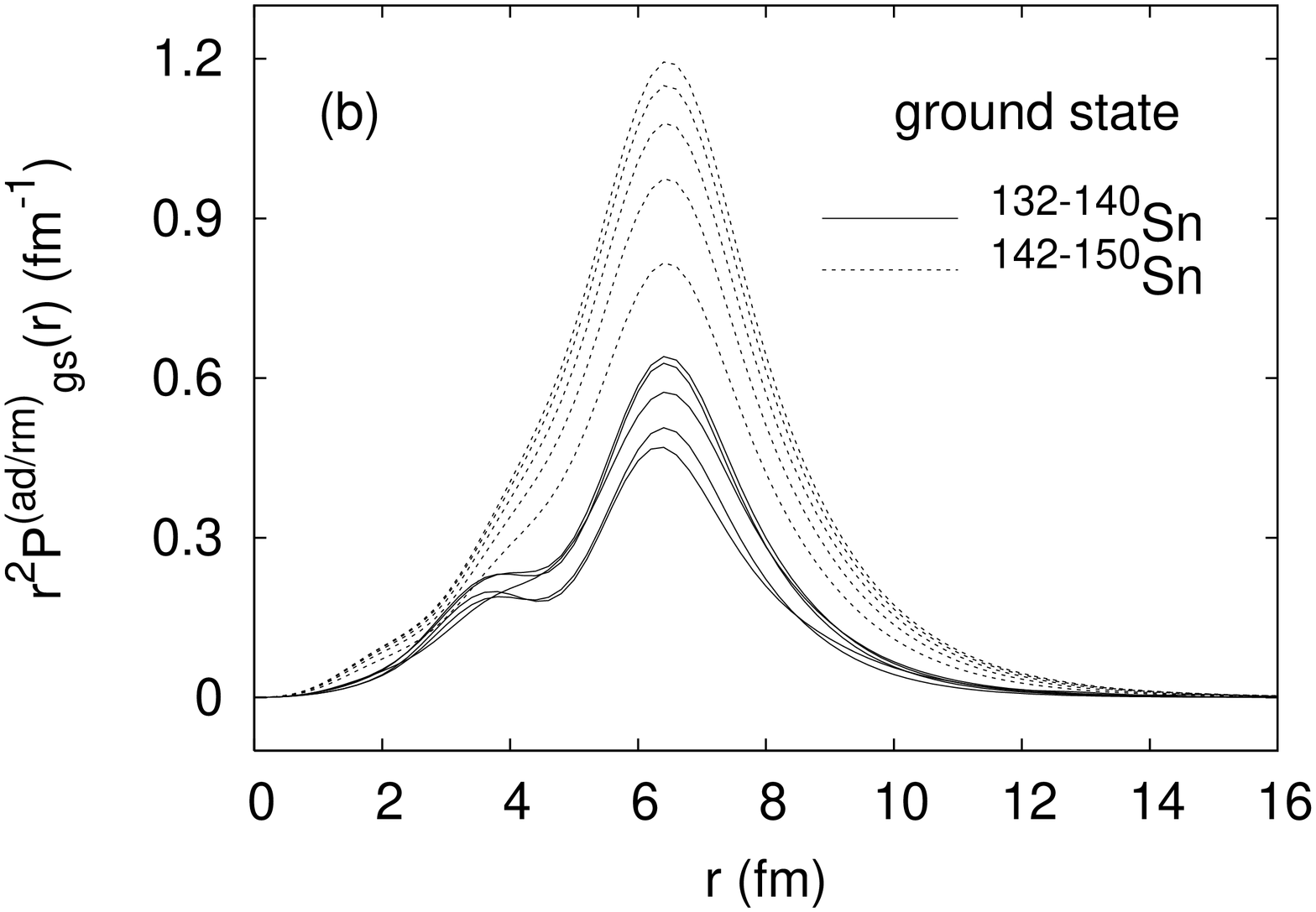}
\caption{\label{fig2}
(a) The neutron pair transition density $r^2P^{({\rm ad/rm})}_{{\rm gs}L=0}(r)$  
for the ground-state transitions in the
 isotopes $^{120-130}$Sn and  $^{134}$Sn.
(b) The same as (a), but in $^{132-140}$Sn and $^{142-150}$Sn.
The transition density for $^{132}$Sn is the pair-addition transition
density $r^2P^{({\rm ad})}_{{\rm gs}L=0}(r)$
calculated in the QRPA.
}
\end{figure*}

Let us first discuss the 
pairing rotation, i.e., the monopole transition with $L=0$ from the ground state of a superfluid open-shell 
even-$N$ isotopes
 to the ground state of neighboring $N\pm 2$ isotopes. 
The transition density and the transition matrix element can be
calculated at the level of the static HFB as
 the transition density for this transition
can be approximated as
\begin{equation}
P_{{\rm gs}}^{({\rm ad/rm})}(\vecr)\equiv
\langle \Psi_{0,N\pm 2} |
\psi^\dagger(\vecr\downarrow)\psi^\dagger(\vecr\uparrow)| \Psi_{0}\rangle
\approx  \frac{1}{2} \tilde{\rho}(\vecr)
\end{equation}
where $ \tilde{\rho}(\vecr)$ is the pair density defined by
\begin{equation}
 \tilde{\rho}(\vecr) \equiv  \langle \Psi_{0} |
\sum_{\sigma}\psi^\dagger(\vecr\sigma)\psi^\dagger(\vecr\tilde{\sigma})| \Psi_{0}\rangle.
\end{equation}
The radial transition density is given as
\begin{eqnarray}\label{gs-transition-density}
P_{{\rm gs} L=0}^{({\rm ad/rm})}(r)~=~\sqrt{\pi}\tilde{\rho}(r).
\end{eqnarray}
In this approximation, there is no distinction between the addition and removal modes.
The strength 
of the ground-state transfer is calculated  as
\begin{equation}\label{gs-strength}
B({\rm Pad/rm}0; {\rm gs} \rightarrow {\rm gs})
=
\left|\sqrt{\pi}\int r^{2}\tilde{\rho}(r)dr\right|^2.
\end{equation}
Note that in
the closed-shell isotopes $^{100}$Sn and $^{132}$Sn
we calculate the strength and the transition density in the
QRPA instead of Eqs.~(\ref{gs-transition-density}) and (\ref{gs-strength}) which 
are not appropriate to nuclei with vanishing pairing gap. 

Figure~\ref{fig1}(a) is the plot of  the calculated strength $B({\rm Pad/rm}0)$ of the ground-state
two-neutron transfer. The strength is enhanced significantly as the mass number
(the neutron number) exceeds $A=140$ ($N=90$).  
For the isotopes with $A>140$, the absolute magnitude
of the pair transfer strength reaches more than twice the maximum strength
in the  region $100<A <132$.
In the conventional BCS approximation,
the pair transfer strength is proportional to $(\Delta/G)^2 \propto \Delta^2$ where
$\Delta$ is the pairing gap and $G$ is the force strength of the seniority pairing force
\cite{Yoshida62,Broglia73,Brink-Broglia}.
In Fig.~\ref{fig1}(b), we plot the square $\Delta_{uv}^2$ of the average pairing gap
\begin{equation}
\Delta_{uv}=
\frac{\int \rhot(\vecr)\Delta (\vecr)d\vecr}
{\int\rhot(\vecr)d\vecr}\\
\end{equation} 
of neutrons.
Comparing the isotopic trends of the two quantities, we see
relative enhancement of the two-neutron transfer strength by a factor of
two in
$A>140$, and also in $132<A<140$. For the latter isotopes, 
 the absolute magnitude  of the
average pairing gap is small, but the two-neutron transfer
strength is comparable to that of more stable isotopes in
the region $100<A <132$. 

The origin of the enhancement is clarified by looking at the pair
transition density, which is shown in Fig.~\ref{fig2}(a) and (b) for
$120 \le A \le 132$ and $132< A \le 150$, respectively. It is seen that  the profile of
the transition
density suddenly changes as the
neutron number exceeds the $N=82$ magic number and $N=90$. The transition
density for $132< A \le 150$ extends outside the surface,
reaching  $r\sim 11$ fm for $132 <A <140$, 
and $r \sim 14$ fm for  $140 <A <150$.
The amplitude in the exterior region $ r \gesim 7$ fm for $A\geq 132$
is evidently larger than those for $A <132$, where the amplitudes extend only 
up to $r\sim 9$ fm. 
Comparing the results 
for $A=120$ and for $A=144$, for instance,
the maximum values of the amplitude around the nuclear surface
$r \sim 6$ fm are approximately the same, but because of 
 the large spatial extension of the transition
density, the pair transfer strength in $^{144}$Sn is larger by a factor
of $\sim$ 2 (cf. Fig.\ref{fig1}).

The reason for the spatial extension of the pair transition density to develop suddenly  beyond
$N=82$ and $N=90$ can be ascribed to the shell  gap at $N=82$ and 
properties of the neutron single-particle states.
We here note that the transition density of the pair rotational mode, i.e.,
the pair density $\tilde{\rho}(r)$  is written as a coherent sum of
contributions of quasiparticle states, and the
quasiparticle states with lower excitation energy (i.e., those originating from
orbits
close to the Fermi energy) have larger contributions.
The calculated Hartree-Fock single-particle energies for neutrons
in $^{132}$Sn are $e_{{\rm HF}}=-1.99, -0.25, 0.26$ MeV for the
 $2f_{7/2}$, $3p_{3/2}$, and   $3p_{1/2}$
orbits located above the $N=82$ gap, respectively ( $3p_{1/2}$ is an unbound resonance),
and the  $h_{11/2}$ orbit with $e_{{\rm HF}}=-7.68$ MeV is located below the shell gap. 
For the $132<A<140$ isotopes (where the neutron Fermi energy
is located near the position of  $2f_{7/2}$), the main component of the 
transition density originates from this orbit.
Since the binding energy of  $2f_{7/2}$ is rather small, the tail of its wave function
 extends to outside, leading to the long tail in the pair transition density.
When the neutron number exceeds $N=90$ ($A=140$),  the next single particle orbits
$3p_{3/2}$  and $3p_{1/2}$ give large contribution to the pair density. 
Since these $p$ orbits have very small binding or  are unbound,
the spatial extension further develops in isotopes with $N\geq 90$ ($A \geq140$).

\section{Pairing Vibration}

\subsection{Strength function}

\begin{figure*}[t]
\includegraphics[scale=0.3,angle=-90]{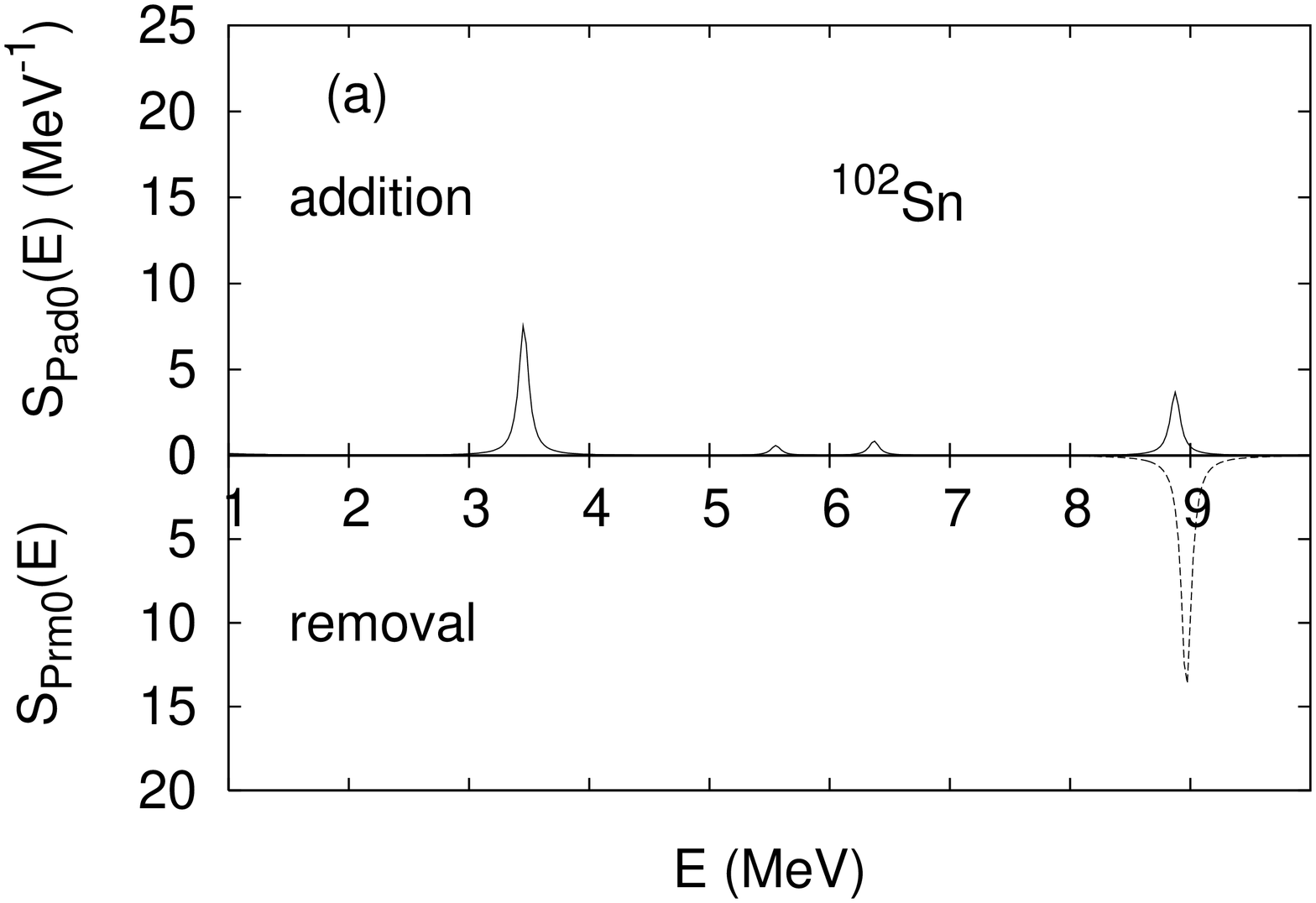}
\includegraphics[scale=0.3,angle=-90]{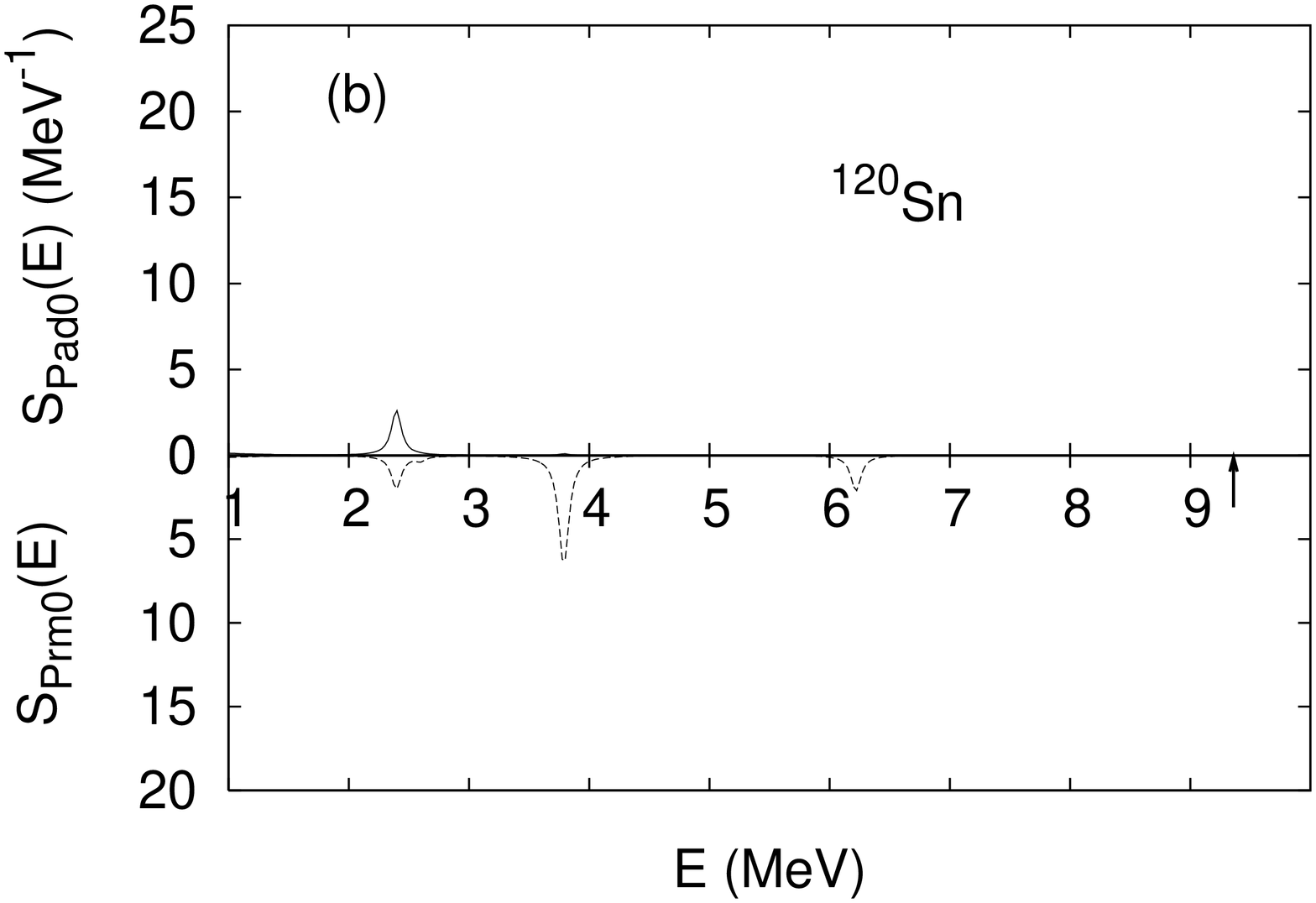}
\includegraphics[scale=0.3,angle=-90]{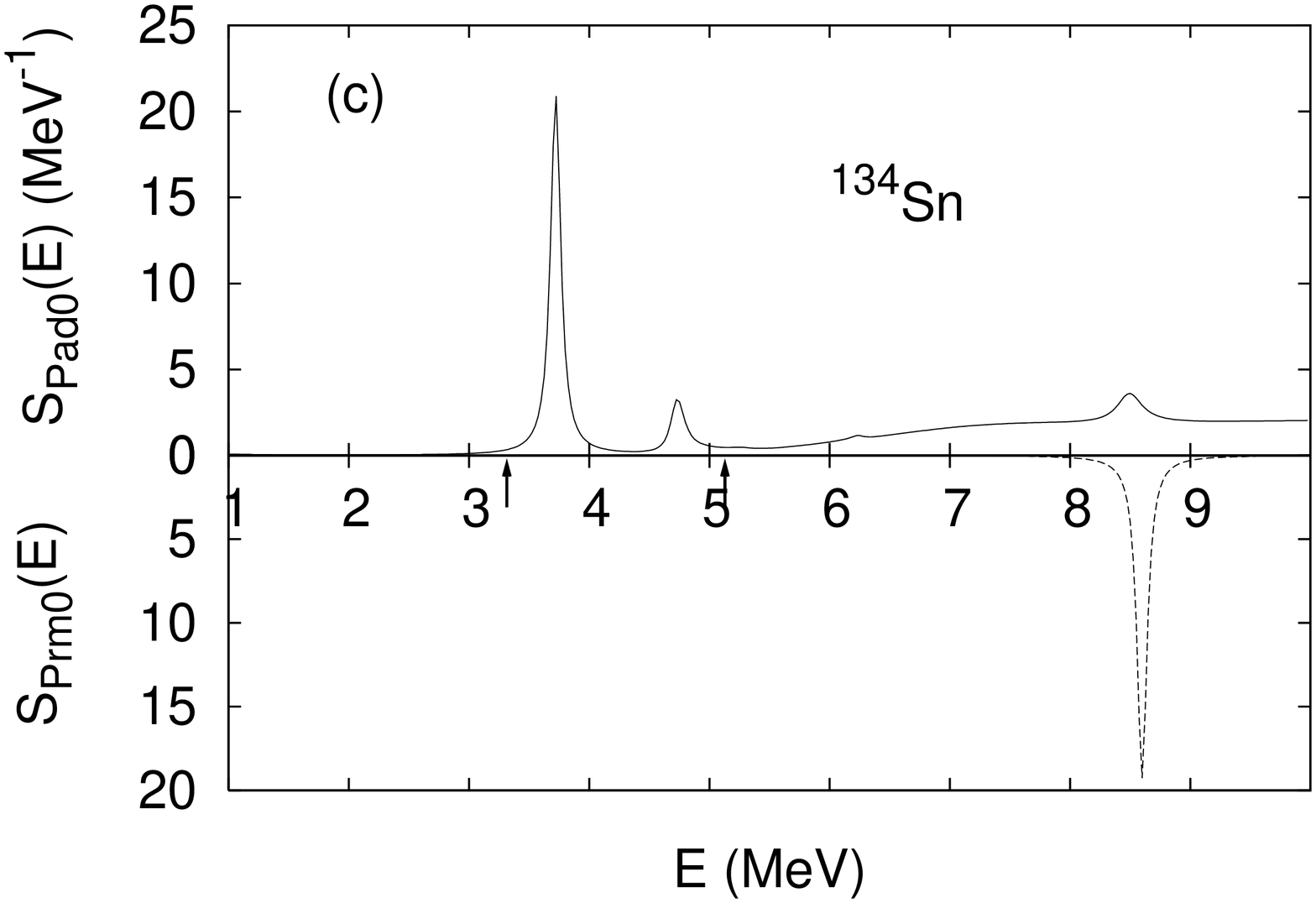}
\includegraphics[scale=0.3,angle=-90]{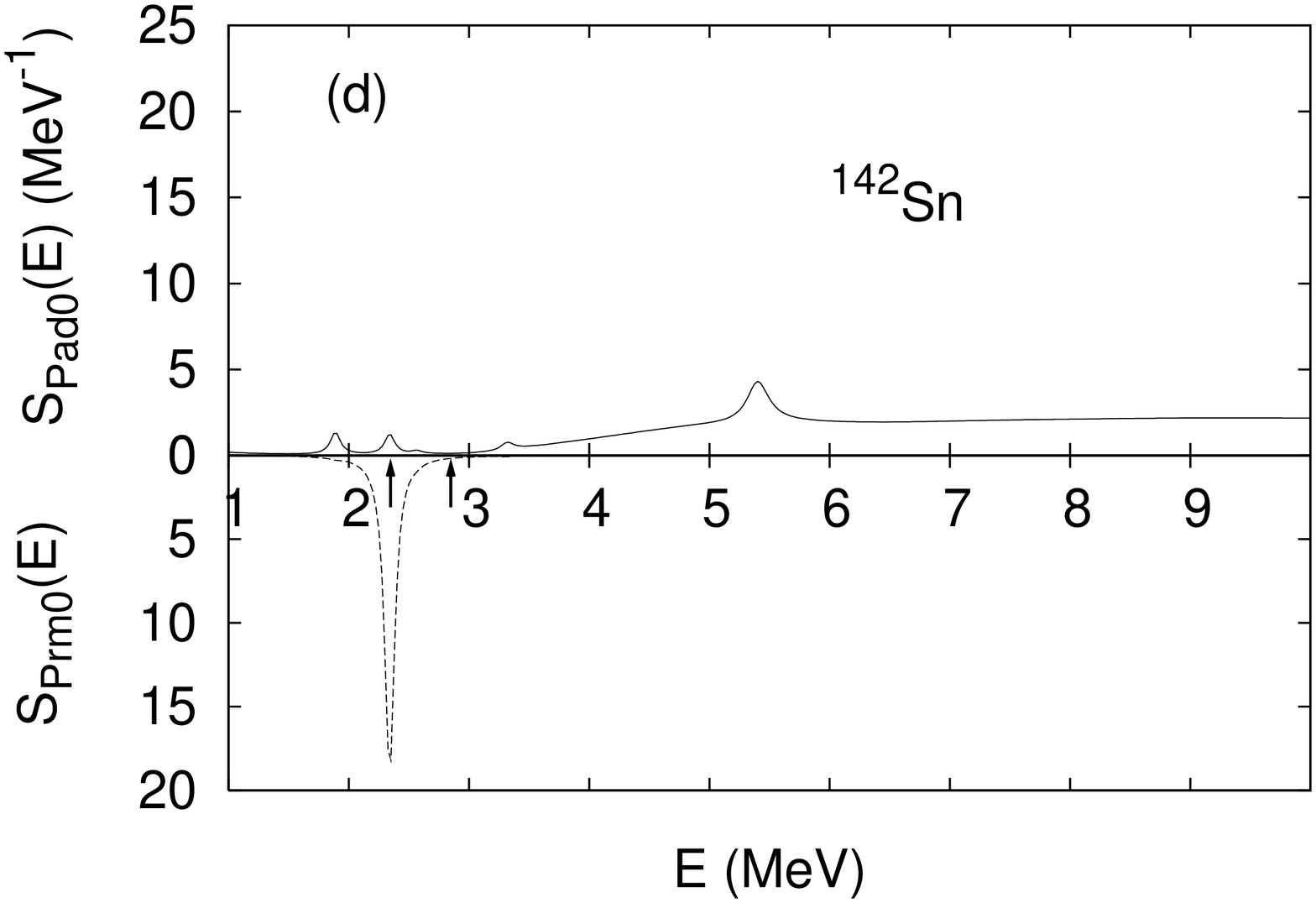}
\caption{\label{fig3}
The strength function $S_{{\rm Pad}0}(E)$ for the
 pair-addition mode (plotted in the upper panel with the solid curve) and the
strength function $S_{{\rm Prm}0}(E)$ for the pair-removal
 mode (the dashed curve in the lower panel) in (a) $^{102}$Sn,
(b)  $^{120}$Sn, (c) $^{134}$Sn and (d) $^{142}$Sn.
The arrows indicate the
 calculated one- and two-neutron separation energies $S_{1n}$ and
 $S_{2n}$.
}
\end{figure*}

We now discuss the two-neutron transfer modes
populating excited  0$^{+}$ states.
Figure \ref{fig3}(a),(b),(c) and (d) show the strength functions
 $S_{{\rm Pad}0}(E)$ and $S_{{\rm Prm}0}(E)$ for the monopole
 pair-addition and -removal modes in four representative isotopes
$^{102}$Sn,$^{120}$Sn,$^{134}$Sn
 and $^{142}$Sn. 
Here, $^{120}$Sn is a mid-shell isotope near the stability line. 
$^{134}$Sn is a representative of  neutron-rich unstable isotope which
is located just beyond the neutron magic number $N = 82$ while $^{102}$Sn 
has the same closed-shell plus two-particle configuration, but at the neutron-deficient side. 
In the most neutron-rich isotope $^{142}$Sn, the neutron Fermi energy is located around
the weakly bound $3p_{3/2}$ orbit. 

All the four isotopes exhibit low-lying peaks 
in the excitation energy range of  $2 \lesim E \lesim 4$ MeV,
and these peaks may be regarded as the pairing vibrational modes.
They corresponds to the lowest QRPA solutions 
 in the cases of $^{102}$Sn, $^{120}$Sn and $^{134}$Sn, 
but 
second lowest solution in the case of
 $^{142}$Sn. 
 The excitation energies of the pair vibrational states are
  3.55, 2.40, 3.73 and 2.35 MeV in 
   $^{102}$Sn, $^{120}$Sn, $^{134}$Sn and $^{142}$Sn, respectively.
They are not very different from  the typical energy
 $E \approx 2\Delta$ ($\sim 2$MeV)
expected
 in the standard pair vibrational model\cite{Bes-Broglia66,Broglia73,Brink-Broglia}.

The most prominent feature seen in Fig.\ref{fig3} is that 
the pair-addition strength 
associated with the low-lying pair vibrational mode
in $^{134}$Sn is several times larger  than
those in the other cases. 
 The pair-addition strength of the pairing vibration is 
 $B({\rm Pad}0)=3.16$ in $^{134}$Sn while it is 
1.18 and 0.40 in $^{102}$Sn and $^{120}$Sn. The magnitude of the
pair addition strength in $^{134}$Sn is so large that it is comparable with 
the strength  $B({\rm Pad/rm}0)=3.61$ associated with the ground-state transition
(the pairing rotation). 
The pair-removal strength is negligibly small, and hence it is
essentially a pure pair-addition mode.
We also notice that the pair vibrational mode in $^{134}$Sn (at $E=3.73$ MeV) 
 is located above the one-neutron separation energy $S_{1n}=3.31$  MeV (indicated with
an arrow in the figure).   We estimate the resonance width
being less than 1 keV by evaluating the  FWHM of the peak minus
 the smoothing width 
$2\epsilon=100$ keV. 
The very small width indicates that the pair vibrational state is a {\it narrow resonance} 
even though it is embedded in the continuum. The above characteristics indicate clearly that
the pair vibrational mode in $^{134}$Sn
deviates from the conventional picture of the pairing vibration.

The pair vibrational mode in $^{142}$Sn has a character different from that in $^{134}$Sn
as it has the large pair-removal strength instead of the pair-addition strength.

Looking at the strength in $^{134}$Sn and $^{142}$Sn  
at higher excitation energies, it is seen that
there exists a smooth distribution of the 
pair-addition strength above
the two-neutron separation energy $S_{2n}$.
 At any excitation energy above $S_{2n}$, it is
always possible to put two neutrons outside the nucleus  although
these two neutrons immediately escape out of the initial position.
The smooth distribution can be ascribed 
to this process. Since it does not correspond to "transfer" reaction, we do not analyze 
below.  Concerning the pair-removal strength function, sharp and large peaks located
around $E=9.1$ and 8.6 MeV are observed in Fig.~\ref{fig3}(a) and (c), respectively. 
This is the so called
 giant pairing vibration\cite{Broglia-Bes77,Herzog85,Lotti89,Oertzen-Vitturi}:
it is
 a collective mode of removing two neutrons which
occupy the single-particle orbits in the next major-shell below the valence shell, 
i.e. the orbits in
the $N=50-82$ shell in the case of $^{134}$Sn. 
 In the cases of $^{120}$Sn and $^{142}$Sn,
similar peaks exist, but at slightly higher energy,  and are 
not seen in Fig.~\ref{fig3}(b) and (d).

\subsection{Transition density}

\begin{figure*}[t]
\includegraphics[scale=0.3,angle=-90]{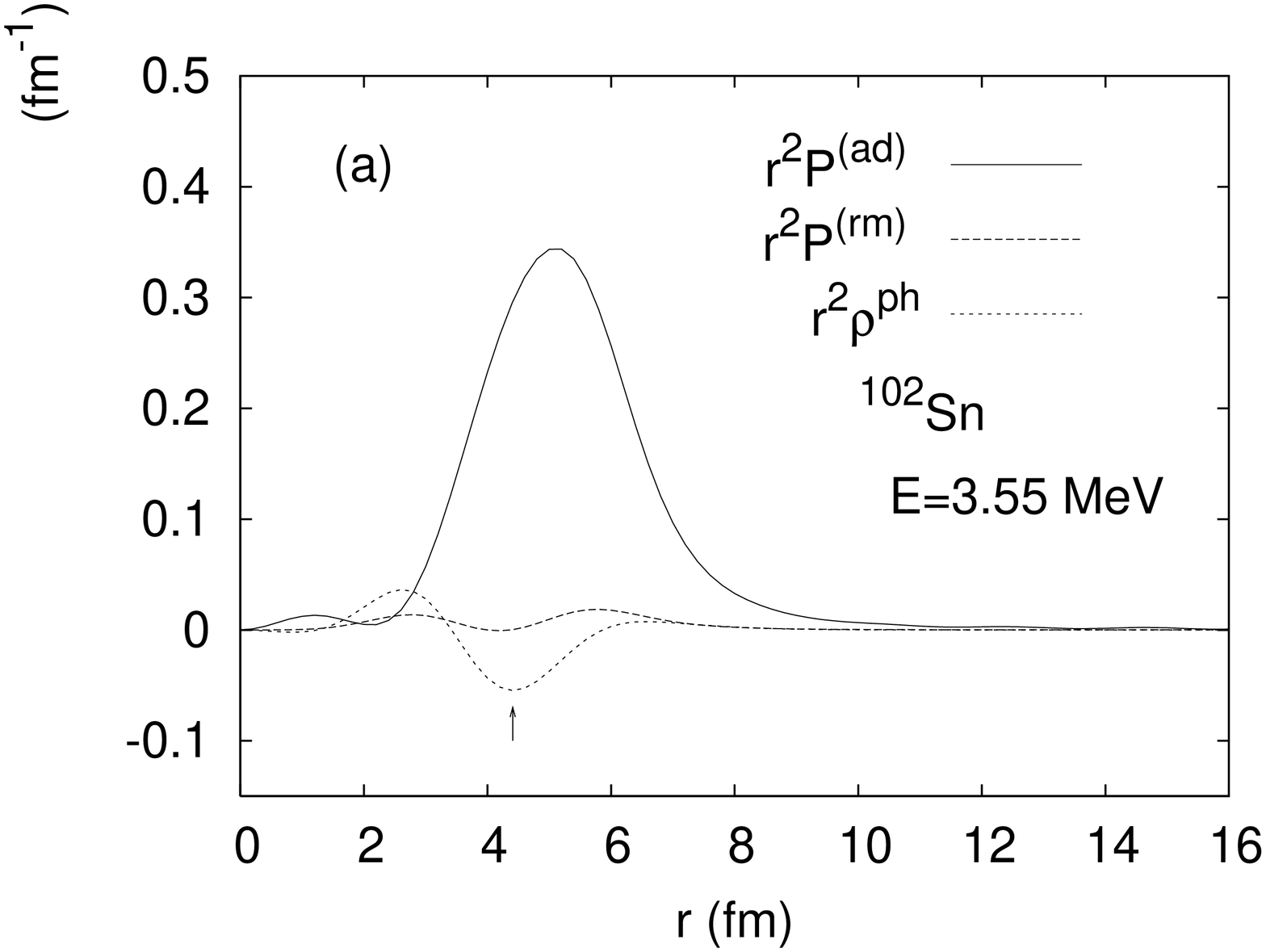}
\includegraphics[scale=0.3,angle=-90]{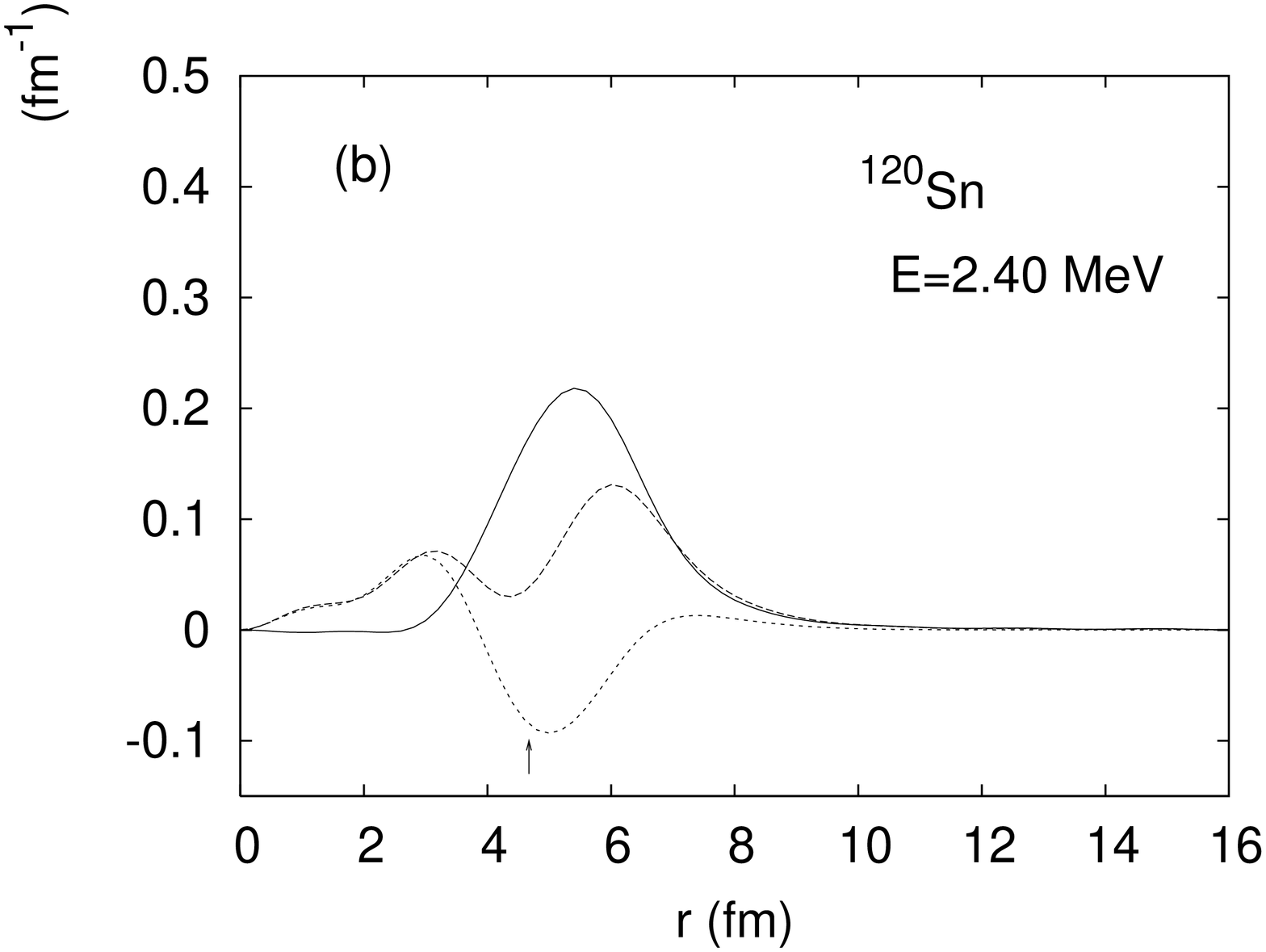}
\includegraphics[scale=0.3,angle=-90]{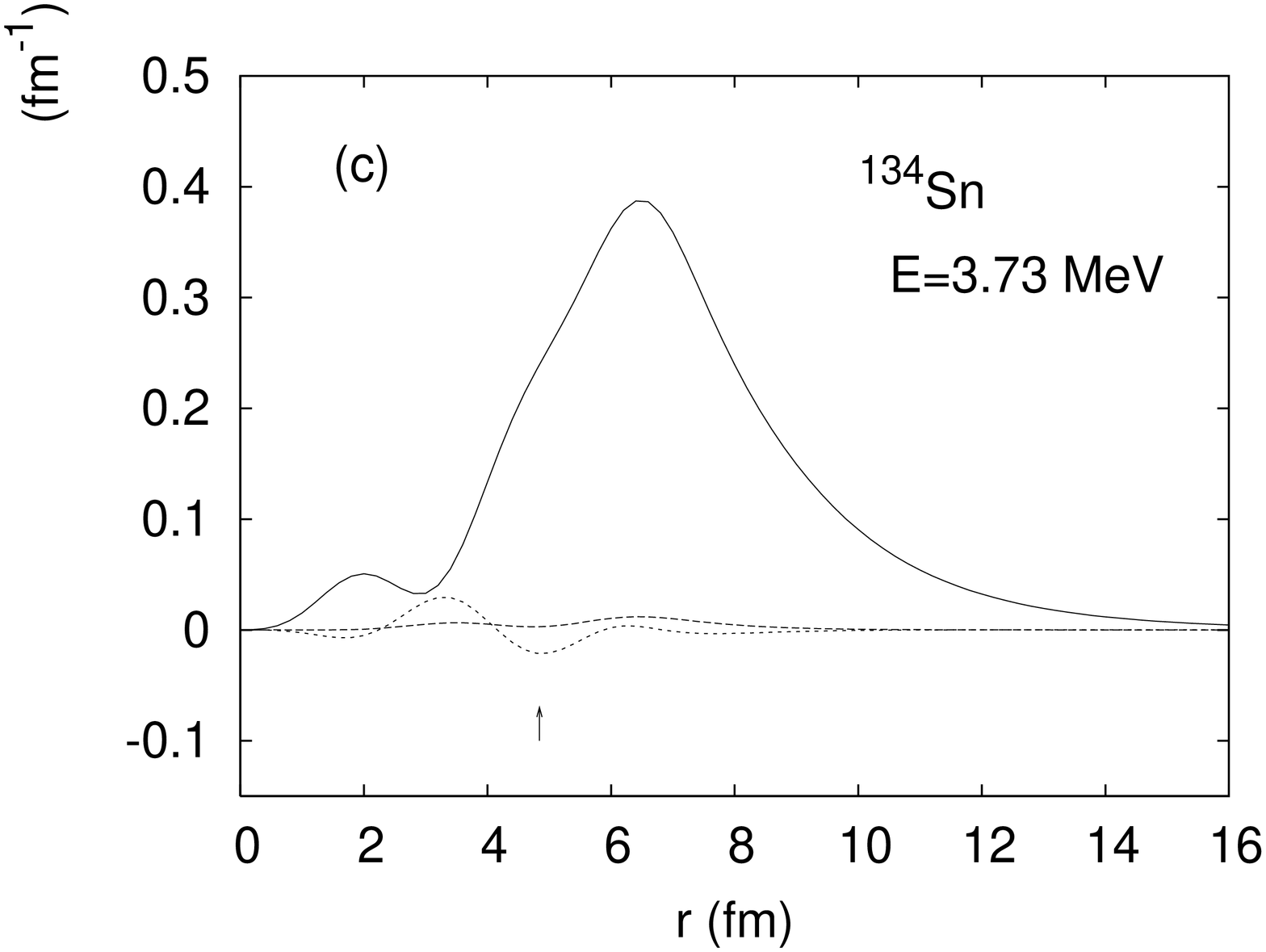}
\includegraphics[scale=0.3,angle=-90]{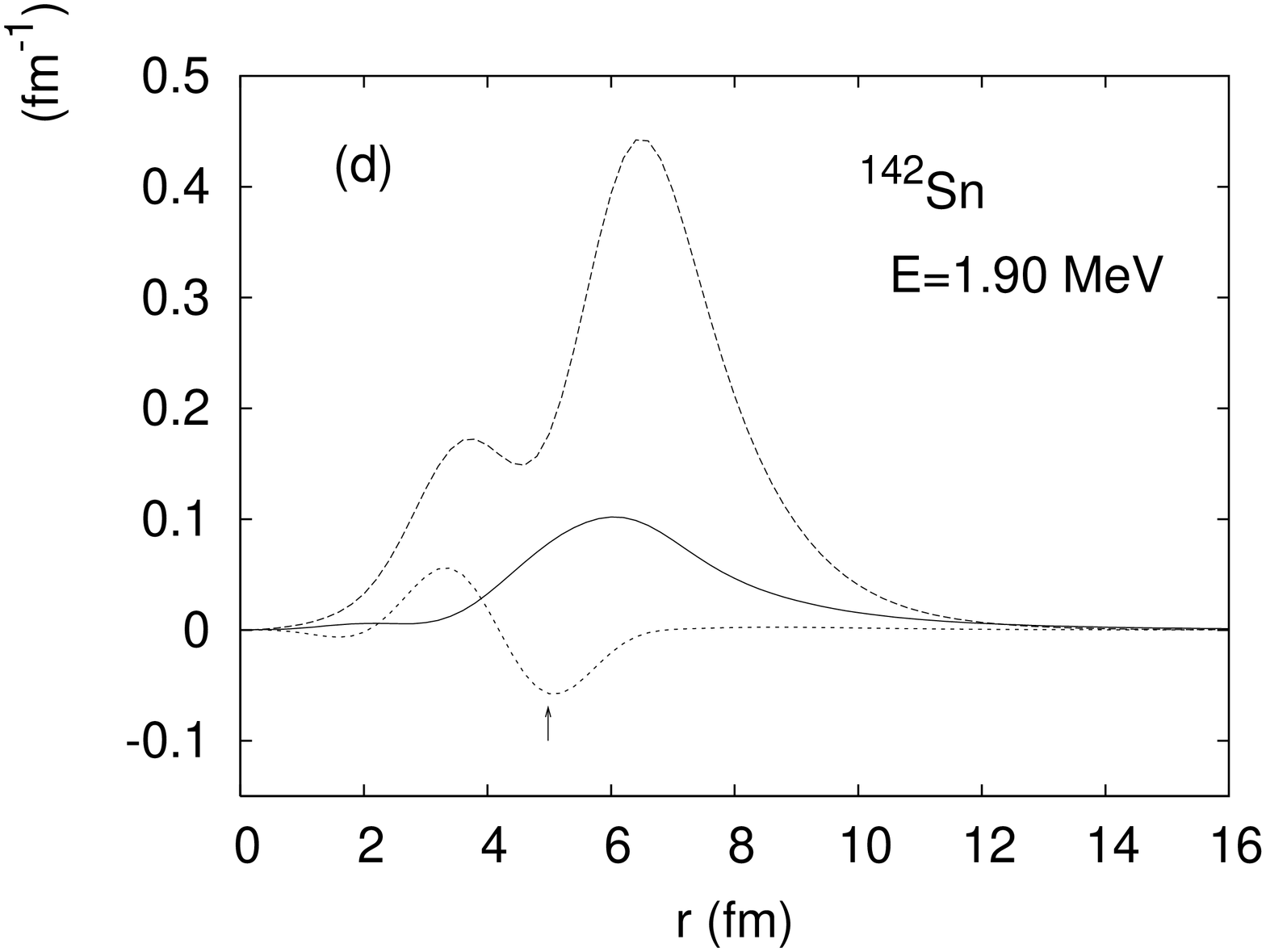}
\caption{\label{fig4}
(a) The neutron transition densities associated with the pair vibrational mode 
at $E=3.55$ MeV in
 $^{102}$Sn. The solid, the dashed  and the dotted curves represent the pair-addition
 transition density $r^{2}P^{({\rm ad})}_{i L=0}(r)$, the pair-removal transition
 density $r^{2}P^{({\rm rm})}_{i L=0}(r)$, and  the particle-hole
 transition density $r^{2}\rho^{{\rm ph}}_{i L=0}(r)$, respectively.
(b)(c)(d) The same as (a) but for $^{120}$Sn, $^{134}$Sn and $^{142}$Sn,
respectively.
The position of the arrow  indicate the
the 
root mean square radius $R_{{\rm rms}}=\sqrt{\left<r^2\right>}$ of the total nucleon density, which 
are
 4.41, 4.67, 4.84 and 4.98 fm in $^{102}$Sn, $^{120}$Sn, $^{134}$Sn and $^{142}$Sn,
respectively.
}
\end{figure*}

We show in
Fig.\ref{fig4} the pair-addition and -removal transition densities 
$P^{({\rm ad})}_{i L=0}(r)$ and $P^{({\rm rm})}_{i L=0}(r)$,
and the particle-hole transition density $\rho^{{\rm ph}}_{i L=0}(r)$ for neutrons, associated with the
low-lying pair vibrational modes 
 in (a) $^{102}$Sn, 
(b)  $^{120}$Sn, (c)  $^{134}$Sn and (d) $^{142}$Sn. 
The volume element $r^2$ is multiplied to these quantities in the plots.

A peculiar feature of the low-lying pair vibrational mode in $^{134}$Sn is clearly visible in 
Fig.\ref{fig4} when it is compared with  the results for $^{102}$Sn  and $^{120}$Sn. In the cases
 of $^{102}$Sn  and $^{120}$Sn, 
all of the three kinds of transition density have increased amplitudes around
the nuclear surface, and the amplitudes  diminish quickly far outside the surface, 
$r \gesim R_{{\rm rms}} + 3$ fm ($r \gesim  8$ fm).
In $^{134}$Sn, however, the pair-addition transition density  $P^{({\rm ad})}_{i L=0}(r)$
has a significant amplitude even at $r = R_{{\rm rms}}+3$ fm ($r \approx 8$ fm)
and a long tail extends up to $r\approx 15$ fm. 
It is interesting to compare  $^{102}$Sn  and $^{134}$Sn (Fig.\ref{fig4}(a) vs. (c))
since  both cases   commonly have the closed-shell plus two-neutron configuration. 
Although the maximum amplitudes in the two cases are about the same, 
the transition density in $^{134}$Sn significantly extends toward outside, resulting
in the noticeable increase of the pair-addition transfer strength in $^{134}$Sn,
i.e., $B({\rm Pad}0)=3.16$, which is larger by a factor of three than $B({\rm Pad}0)=1.18$ in
$^{102}$Sn. We thus find that the spatial extension of the transition amplitude 
plays a central role to enhance the pair-addition transfer strength $B({\rm Pad}0)$
 in $^{134}$Sn.
 
The low-lying pair vibrational mode in $^{142}$Sn exhibits also an extended profile
in the transition density, but in this case
 the dominant amplitude is the pair-removal transition density 
$P^{({\rm rm})}_{i L=0}(r)$
and its spatial extension is smaller (observed up to $r\sim 11$ fm) than 
that of $P^{({\rm ad})}_{i L=0}(r)$ in $^{134}$Sn.

\begin{figure*}[t]
\includegraphics[scale=0.3,angle=-90]{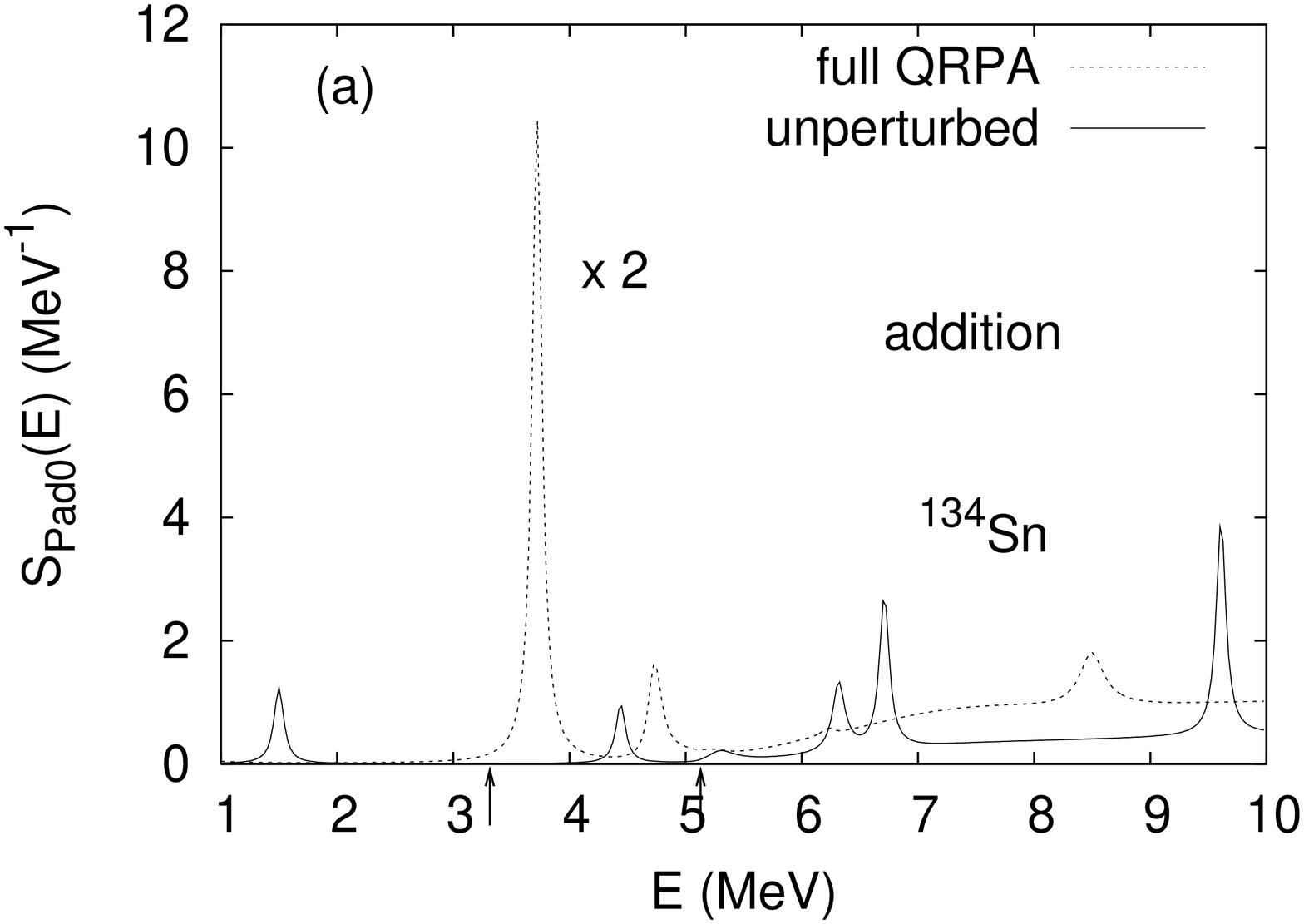}
\includegraphics[scale=0.3,angle=-90]{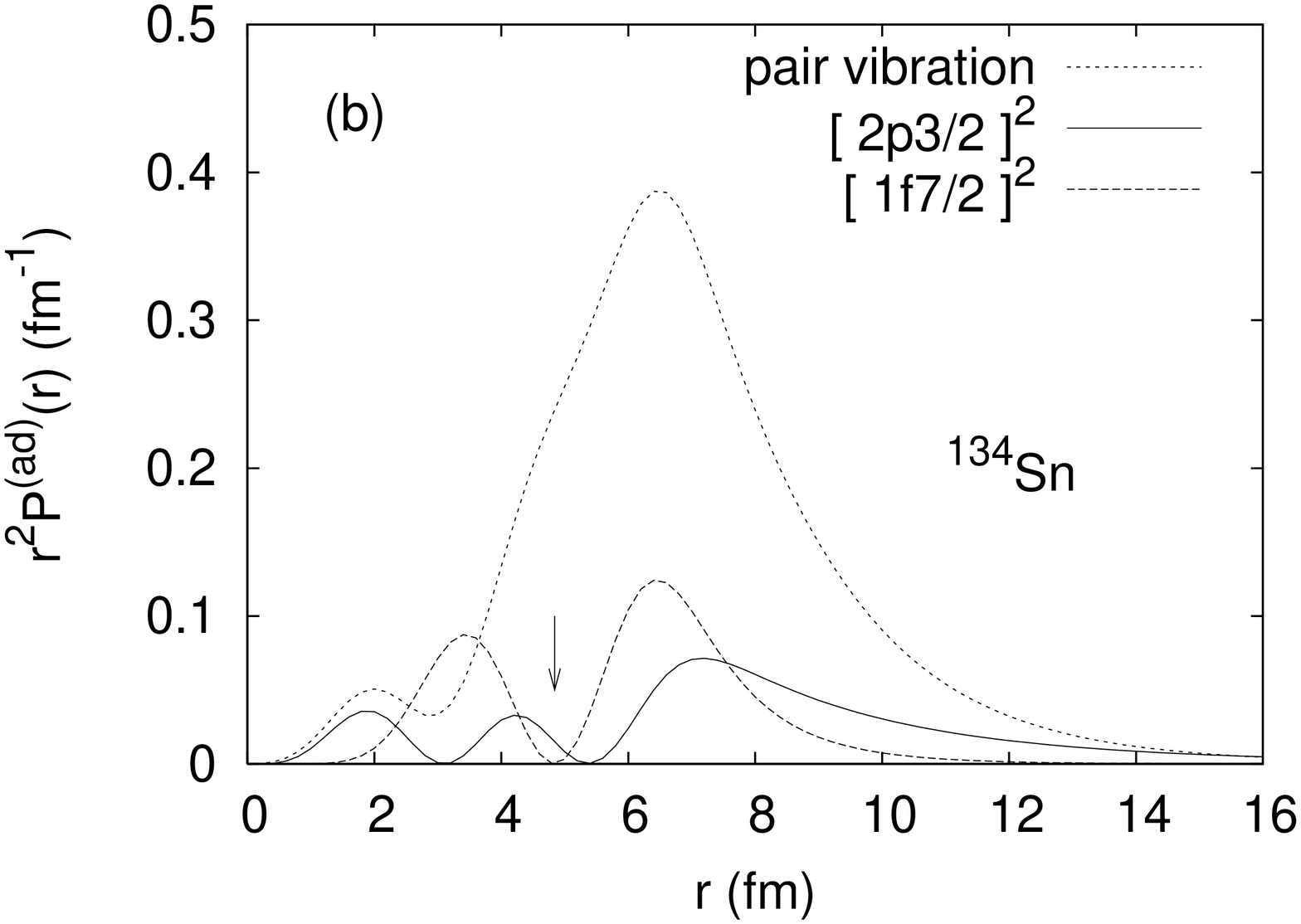}
\caption{\label{fig5}
(a) The pair-addition strength function $S_{{\rm Pad}0}(E)$ in $^{134}$Sn, calculated for
the unperturbed neutron two-quasiparticle excitations (plotted with the sold curve) and in
the full QRPA (dotted curve). 
(b) The
 transition density $r^{2}P^{({\rm ad})}_{i L=0}(r)$ of the unperturbed neutron two-quasiparticle
 excitations $[2f_{7/2}]^2$ (solid curve) 
and  $[3p_{3/2}]^2$ (dashed curve) in $^{134}$Sn, compared with the
transition density of the low-lying pair vibrational mode
obtained in the full QRPA (dotted curve).
}
\end{figure*}

\begin{figure*}[t]
\includegraphics[scale=0.3,angle=-90]{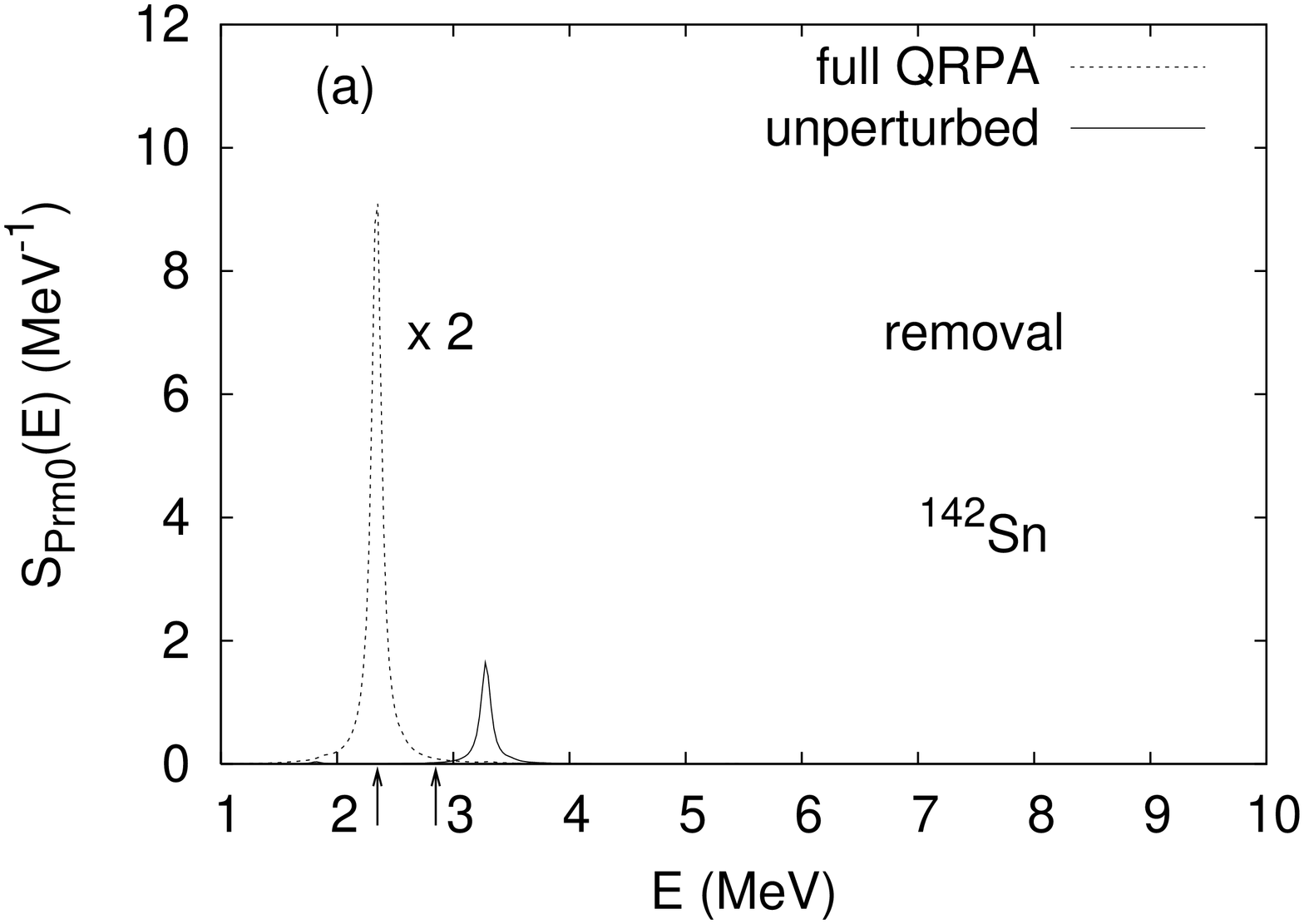}
\includegraphics[scale=0.3,angle=-90]{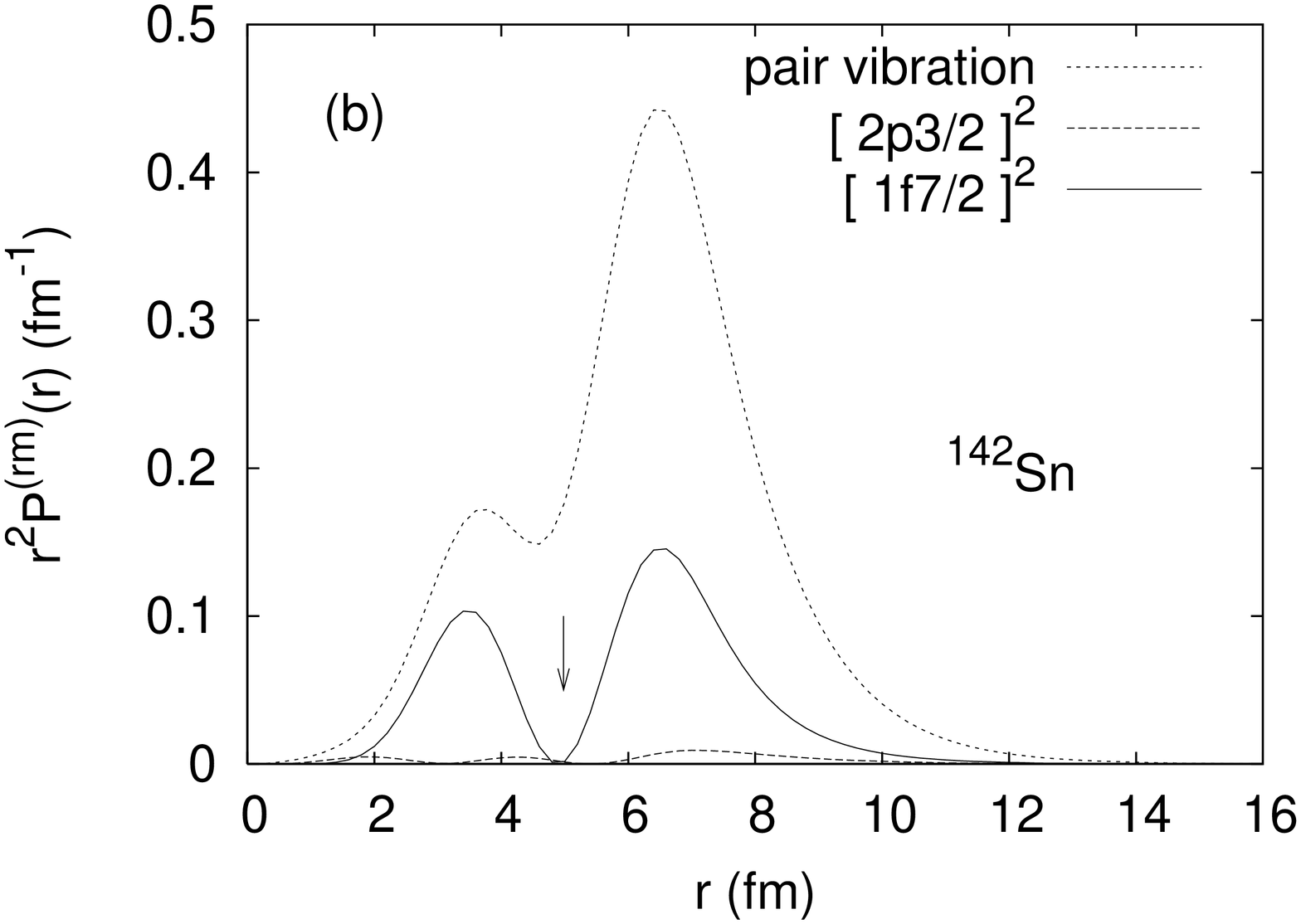}
\caption{\label{fig6}
(a) The pair-removal strength function $S_{{\rm Prm}0}(E)$ in $^{142}$Sn, calculated for
the unperturbed neutron two-quasiparticle excitations (solid curve) and in
the full QRPA (dotted curve).
(b) The
 transition density $r^{2}P^{({\rm rm})}_{i L=0}(r)$ of the unperturbed neutron two-quasiparticle
 excitations $[2f_{7/2}]^2$ (solid curve) 
and  $[3p_{3/2}]^2$ (dashed curve)  in $^{142}$Sn, compared with the
transition density of the low-lying pair vibrational mode
obtained in the full QRPA (dotted curve).
}
\end{figure*}

\subsection{Microscopic origin}

In order to clarify the microscopic origin of the
anomalous pair vibrational mode  in $^{134}$Sn, we show in
Fig.~\ref{fig5}(a) effects of the residual pairing interaction on the pair-addition strength
function $S_{{\rm Pad}0}(E)$. Plotted 
here is the strength function associated with unperturbed neutron two-quasiparticle
excitations, i.e., the strength obtained by neglecting the residual interaction,
and it is compared with the result of the full QRPA calculation.
The first peak at $E\approx 1.5$MeV and the second one at $E\approx 4.5$ MeV correspond
to the two-quasineutron excitations $[2f_{7/2}]^2$ and  $[3p_{3/2}]^2$, respectively.
The strengths of these peaks are $B({\rm Pad}0)=0.18$ and $0.15$ for 
 $[2f_{7/2}]^2$ and  $[3p_{3/2}]^2$, respectively, and they are comparable to a single-particle
 estimate $B_{{\rm s.p.}}({\rm Pad}0)=(2j+1)/8\pi$.
The large low-lying strength does not show up in the unperturbed strength, indicating that the collective
configuration mixing caused by the residual interaction plays a crucial role to bring
about the large strength of the pair vibrational mode in the full QRPA calculation. 
The collective enhancement
 is a factor of twenty
 as is
estimated from the ratio of the strengths (3.16 vs. 0.18, 0.15). 

Figure~\ref{fig5}(b) shows
the pair-addition transition densities $P^{({\rm ad})}_{i L=0}(r)$ associated with  unperturbed
neutron two-quasiparticle excitations $[2f_{7/2}]^2$ and  $[3p_{3/2}]^2$.
It is seen that the
transition density of the $[3p_{3/2}]^2$ configuration has a long tail, extending up to
$r\sim 15$ fm. 
 We here note that the 
quasiparticle energy of the neutron  $2f_{7/2}$ and  $3p_{3/2}$ states in
$^{134}$Sn are 0.75 and 2.25 MeV, and the energies of the corresponding
Hartree-Fock single-particle orbits are
$e_{{\rm HF}}=-2.14$ and $-0.36$ MeV, respectively. 
Here the binding energy of the
$3p_{3/2}$ orbit is only one fifth of
that of $2f_{7/2}$, and is smaller than those of 
the single-particle orbits below the $N=82$ gap ($e_{{\rm HF}}< -7.77$ MeV) by a factor of twenty. Clearly  
the long tail associated with the $[3p_{3/2}]^2$ configuration is a consequence of
the weak binding of the $3p_{3/2}$ quasineutron state.

We speculate that many two-quasiparticle configurations
including  $[3p_{3/2}]^2$ and $[2f_{7/2}]^2$
contribute to produce the pair vibrational mode in $^{134}$Sn
as the transition density
is larger  by several times than those of the individual 
unperturbed two-quasiparticle excitations.
Comparing the transition density of  $[3p_{3/2}]^2$ 
with that of the pair vibrational mode, we deduce that the long tail in the 
pair vibrational mode
may be inherited from that of  $[3p_{3/2}]^2$.
Two-quasineutron configurations in the continuum, including the next lying
$3p_{1/2}$ unbound  resonance state (located at positive energies around 0.2 MeV)
may contribute also.  Considering the
orthogonality of the wave function of the pair vibrational mode to that of the
pair rotational mode whose main component is $[2p_{7/2}]^2$,
we deduce that the $[3p_{3/2}]^2$  and  $[3p_{1/2}]^2$ configurations are the largest components
in the pair vibrational mode.

In Fig.~\ref{fig6}(a), we plot the pair-removal strength function arising from 
unperturbed neutron two-quasiparticle excitations in $^{142}$Sn. 
The low-lying peak seen at $E=3.3$ MeV is the 
 $[2f_{7/2}]^2$ two-quasiparticle configurations, corresponding
 to creating two holes (i.e., removing two neutrons) in the $2f_{7/2}$ orbit. 
There are also peaks
 associated with  the unperturbed $[3p_{3/2}]^2$ and $[3p_{1/2}]^2$
 configurations, which are however hardly visible in this scale.
It is seen that  the strength
 of the pair vibrational mode in the full QRPA calculation is
 ten times larger than the unperturbed strength, indicating a collective
 effect. The transition densities of the two-quasiparticle configurations
 $[2f_{7/2}]^2$ and  $[3p_{3/2}]^2$ are shown in Fig.~\ref{fig6}(b).  
Comparison of the full and the unperturbed transition densities suggests 
that $[2f_{7/2}]^2$
is one of the main configurations, but a significant  mixing
 of configurations other than $[2f_{7/2}]^2$  is also present.
 Since it is the pair-removal mode
the configuration mixing involving  those below the $N=82$ shell gap can contribute, but 
the weakly bound and continuum orbits such as  $3p_{3/2}$ and
 $3p_{1/2}$ contribute very little to the pair removal mode as their occupation
are small.
Consequently the spatial extension
of the transition density of  
 the pair-removal vibrational mode in $^{142}$Sn is smaller
than that of the pair-addition vibrational mode in $^{134}$Sn.

\subsection{Systematics}

\begin{figure*}[t]
\includegraphics[scale=0.3,angle=-90]{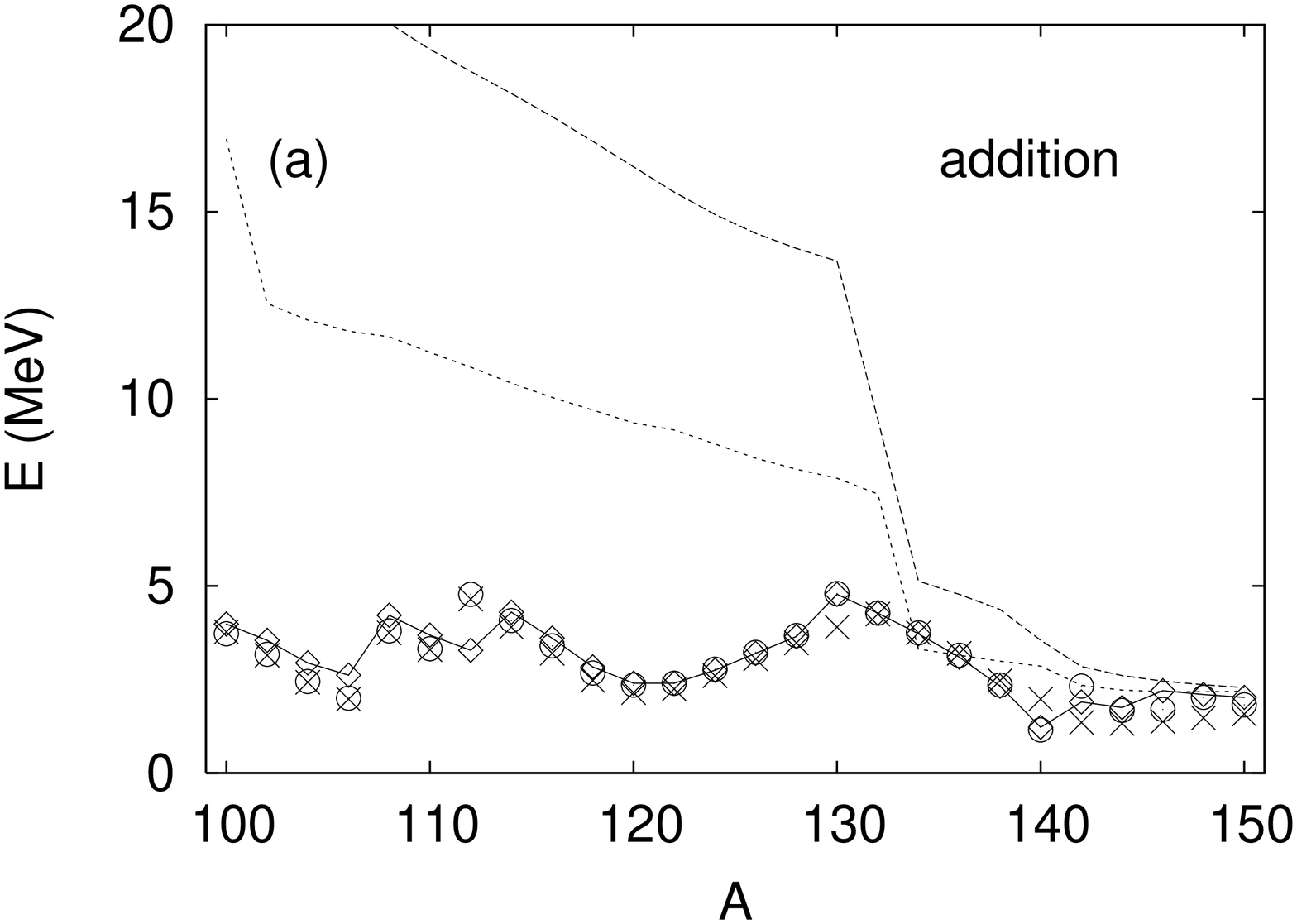}
\includegraphics[scale=0.3,angle=-90]{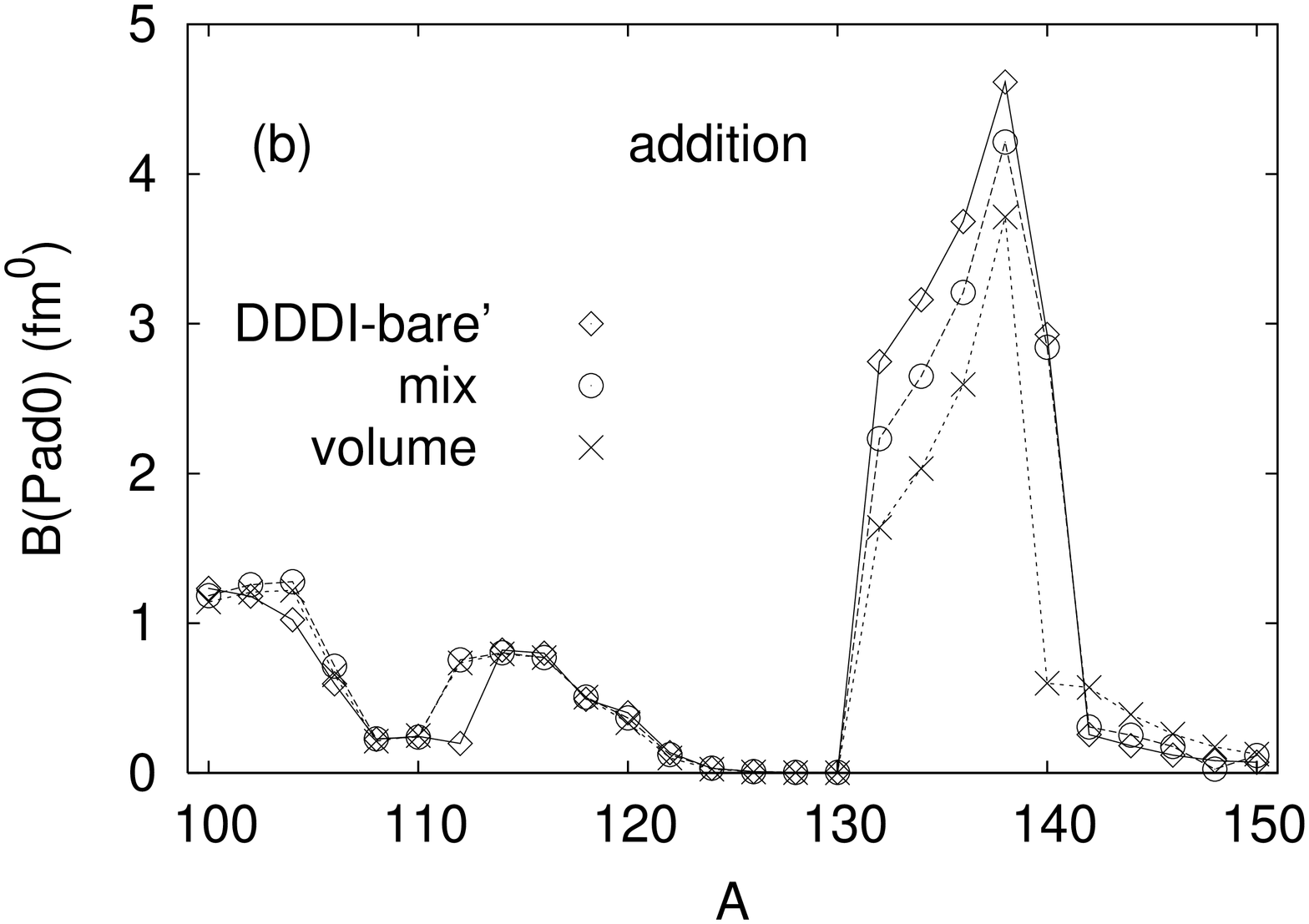}
\caption{\label{fig7}
(a)The excitation energy of the pair-addition vibrational mode.  
 The one- and
 two-neutron separation energies are plotted with the dotted and the
 dashed lines, respectively. 
(b) The pair-addition strength $B({\rm Pad}0)$ of the pair-addition vibrational mode.
 The diamonds connected with the
 solid line are the results obtained with the pairing interaction DDDI-bare' while 
the circles and the crosses are those with the mix and
the volume pairing interactions, respectively.
}
\end{figure*}

\begin{figure*}[t]
\includegraphics[scale=0.3,angle=-90]{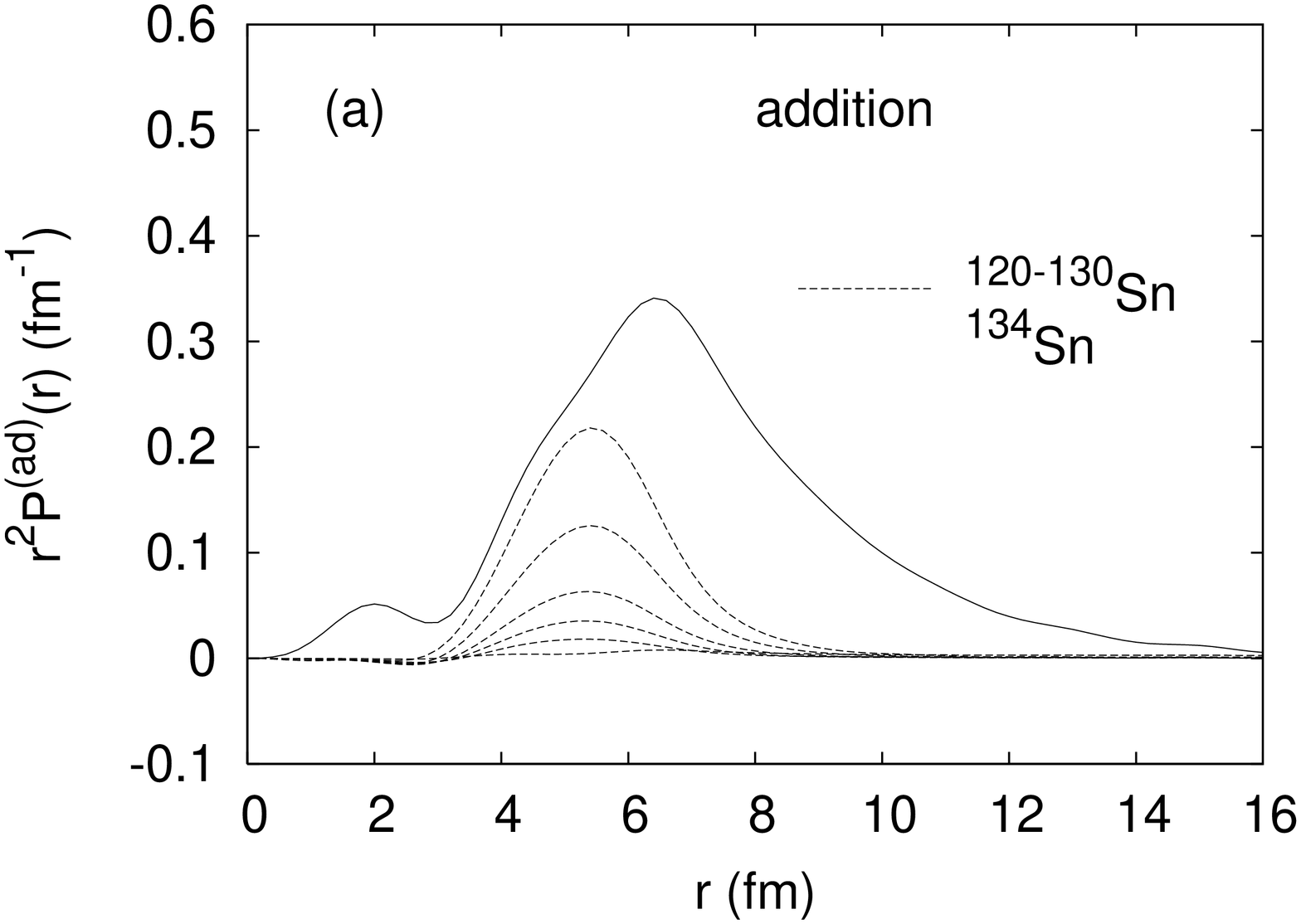}
\includegraphics[scale=0.3,angle=-90]{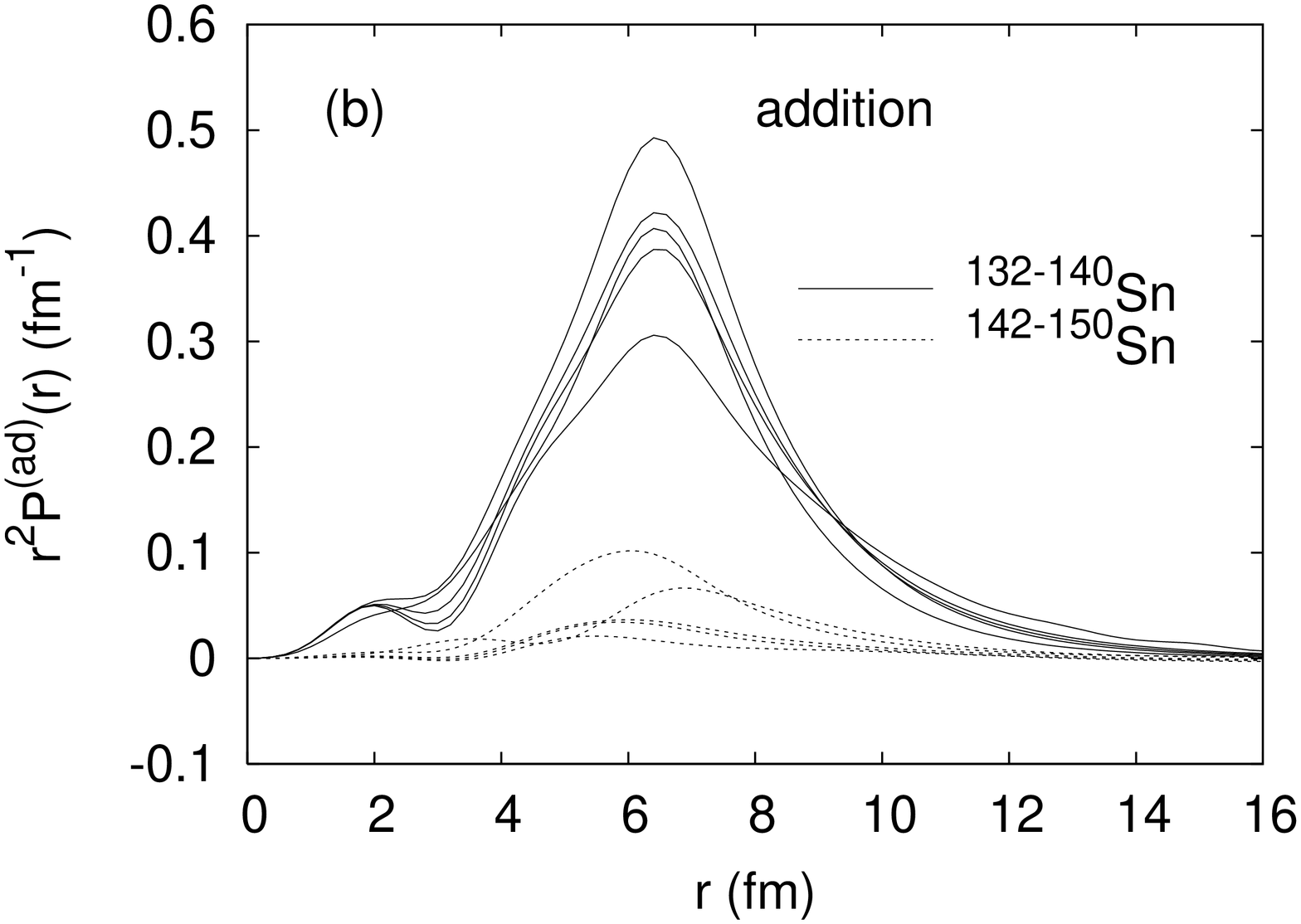}
\caption{\label{fig8}
The pair-addition transition density $r^{2}P^{({\rm ad})}_{i L=0}(r)$ 
associated with the  pair-addition vibrational mode
(a) in $^{120-130}$Sn and $^{134}$Sn and
(b) in $^{132-140}$Sn and  $^{142-150}$Sn. The effective pairing interaction is
 the DDDI-bare'.}
\end{figure*}

It is interesting to see systematical behavior of the low-lying pair vibrational modes
along the isotopic chain of Sn.  Here we treat separately the pair-addition and 
the pair-removal modes
of the pairing vibration. They are identified
as the QRPA eigenmode having the largest pair-addition [-removal] strength among the
peaks below $E=5$ MeV.

In Fig.~\ref{fig7} and Table~\ref{table1}
shown are the excitation energy and the pair-addition strength
$B({\rm Pad}0)$ of the low-lying pair-addition vibrational modes.
 (We plot here the results obtained with the DDDI-bare',
the mix and the volume pairing interactions, but we shall discuss dependence on
the pair interaction later.) 
We plot in Fig.~\ref{fig7}(a) also
the calculated threshold energies $S_{1n}$ and $S_{2n}$ of one- and two-neutron separations. 
A noticeable feature is that in the isotopes
with $A>132$ 
the low-lying pair vibrational
modes  
are located near the neutron separation energies. This is due to the
sudden decrease of the separation energies in isotopes beyond the magic number $N=82$. 
The minimum at $A=140$ ($N=90$) reflects the
sub-shell closure of the $2f_{7/2}$ orbit. 
In Fig.~\ref{fig7} (b), 
large pair-addition strength $B({\rm Pad}0)$ is seen in five isotopes with 
$A=132-140$ 
($N=82-90$).

Figure~\ref{fig8} shows the pair-addition transition density $P^{({\rm ad})}_{i L=0}(r)$ of the
low-lying pair-addition vibrational modes in the isotopes from $A=120$ (a stable isotope) 
to $A=150$. 
We find that the pair vibrational modes in the $A=132-140$ isotopes have quite similar radial
profiles of the transition density $P^{({\rm ad})}_{i L=0}(r)$, and they all share the common character
concerning  the spatial  
extension reaching $r \sim 15$ fm. 
These features all 
indicate that the anomalous pair vibrational mode 
appears systematically in the region beyond the $N=82$ magic number and up to
$N=90$.  
It is also evident 
 that the anomalous pair vibrational modes in $A=132-140$ are distinct from
those in the $A=120-130$ isotopes  closer to the stability line.

\begin{figure*}[t]
\includegraphics[scale=0.3,angle=-90]{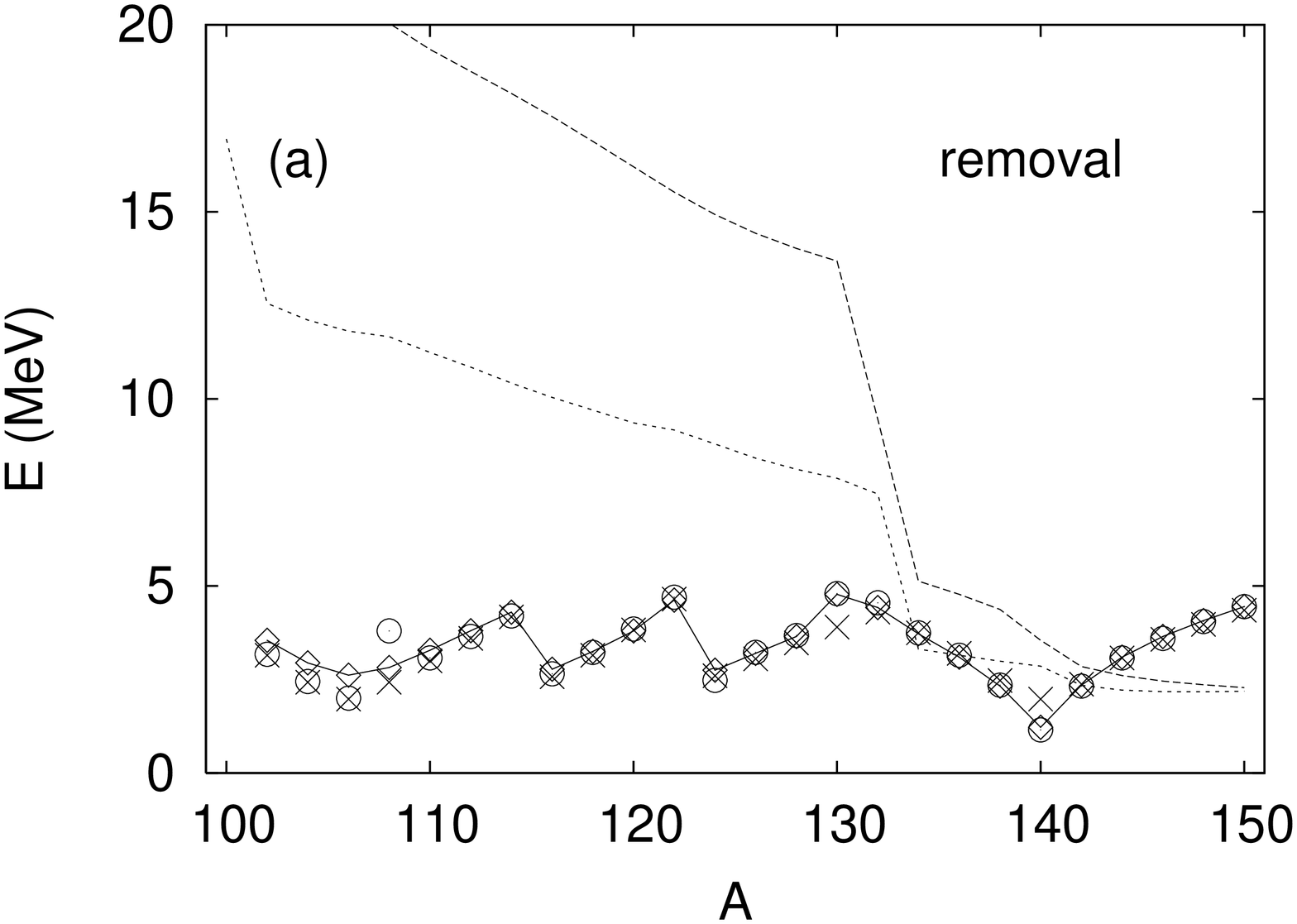}
\includegraphics[scale=0.3,angle=-90]{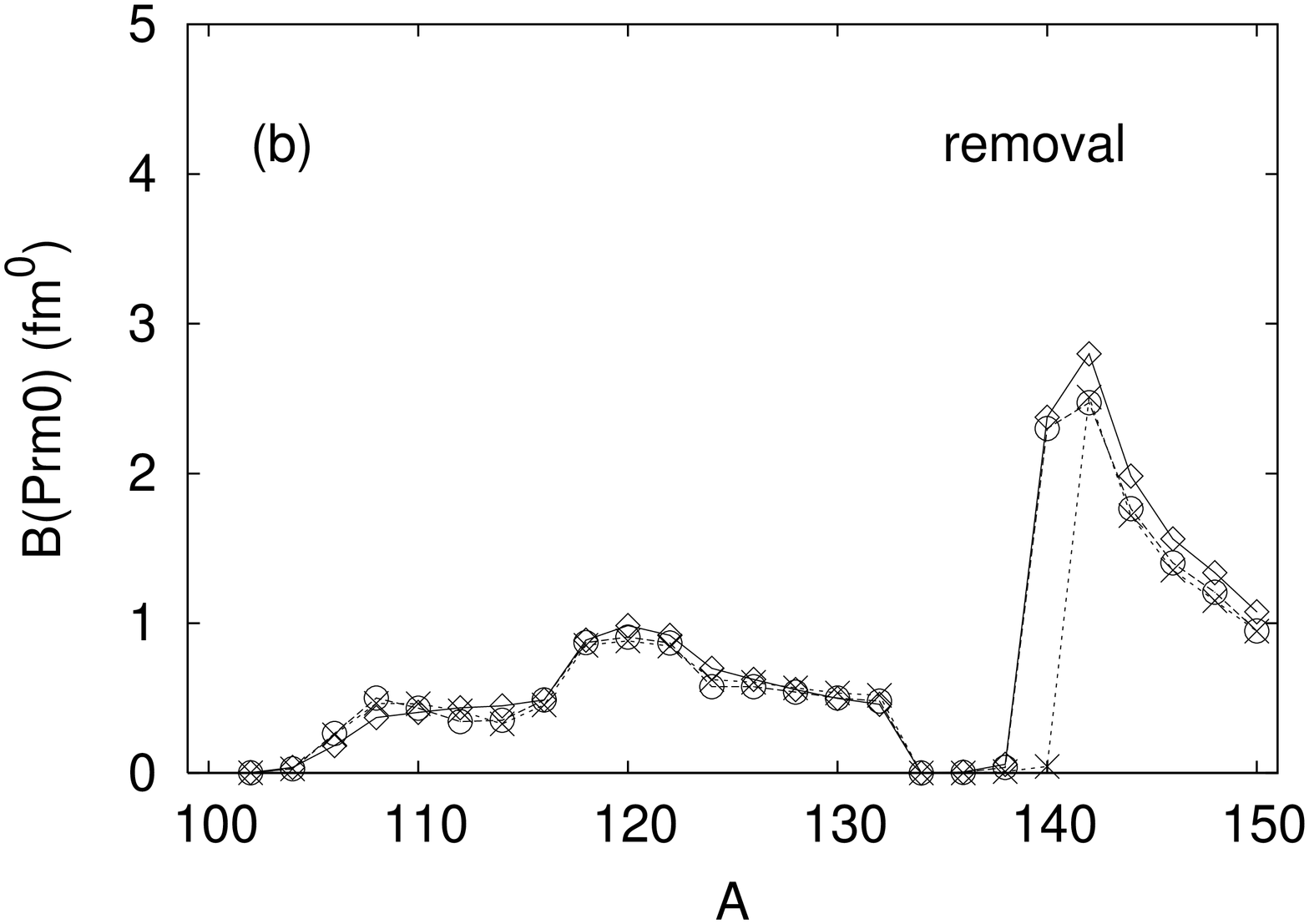}
\caption{\label{fig9}
(a) The excitation energy and
(b) the pair-removal strength $B({\rm Prm}0)$ of the pair-removal vibrational mode.
See also the caption of Fig.~\ref{fig7}.
}
\end{figure*}

Systematics of the pair-removal vibrational mode is shown in
Fig.~\ref{fig9} and also in Table~\ref{table1}.  
The excitation energy and the pair-removal strength $B({\rm Prm}0)$ are 
plotted in Fig.~\ref{fig9}(a)  and (b), respectively. In the isotopes
with $A=134-140$, the low-lying pair-removal strength is negligible.
However,  the pair-removal vibrational mode having large strength emerges in the
isotopes  beyond $A=140$ ($N=90$). It is found that
the pair vibrational mode in these isotopes have the
same character which we discussed 
in connection with the results for $^{142}$Sn (cf. Fig.~\ref{fig4}(d)).

\begin{table}[htb]
\begin{tabular}{cccccc}
\hline
\multicolumn{1}{c}{} 
&\multicolumn{1}{c}{ground state} 
&\multicolumn{2}{c}{p.v. addition}
&\multicolumn{2}{c}{p.v. removal}\\
$A$&  $B({\rm Pad/rm}0)$&  $E$ & $B({\rm Pad}0)$& $E$ & $B({\rm Prm}0)$\\
\hline
100  &   3.096(a)  &  3.99  &  1.233  &        &         \\
102  &   3.138  &  3.55  &  1.180  &  3.55  &  0.003  \\
104  &   5.101  &  2.95  &  1.023  &  2.95  &  0.037  \\
106  &   6.498  &  2.62  &  0.595  &  2.62  &  0.183  \\
108  &   7.857  &  4.22  &  0.225  &  2.82  &  0.370  \\
110  &   8.932  &  3.69  &  0.245  &  3.28  &  0.405  \\
112  &   9.872  &  3.28  &  0.198  &  3.81  &  0.435  \\
114  &   9.798  &  4.30  &  0.820  &  4.30  &  0.448  \\
116  &   9.333  &  3.61  &  0.800  &  2.78  &  0.488  \\
118  &   8.855  &  2.84  &  0.493  &  3.25  &  0.888  \\
120  &   8.274  &  2.40  &  0.402  &  3.80  &  0.984  \\
122  &   7.851  &  2.40  &  0.136  &  4.63  &  0.920  \\
124  &   7.374  &  2.75  &  0.031  &  2.75  &  0.699  \\
126  &   6.462  &  3.20  &  0.010  &  3.20  &  0.625  \\
128  &   4.991  &  3.65  &  0.004  &  3.65  &  0.557  \\
130  &   2.851  &  4.78  &  0.005  &  4.78  &  0.498  \\
132   &   3.202(r) \ 3.973(a)  & 4.28   &  2.747  &  4.43&   0.458  \\
134  &   3.607  &  3.73  &  3.160  &  3.73  &  0.000  \\
136  &   5.662  &  3.10  &  3.683  &  3.10  &  0.007  \\
138  &   6.048  &  2.33  &  4.615  &  2.33  &  0.058  \\
140  &   5.157  &  1.23  &  2.929  &  1.23  &  2.376  \\
142  &  11.079  &  1.90  &  0.256  &  2.35  &  2.800  \\
144  &  16.088  &  1.75  &  0.182  &  3.10  &  1.983  \\
146  &  19.869  &  2.20  &  0.119  &  3.65  &  1.563  \\
148  &  22.639  &  2.10  &  0.082  &  4.08  &  1.338  \\
150  &  24.481  &  2.03  &  0.071  &  4.45  &  1.077  \\
\hline
\end{tabular}
\caption{\label{table1}
The pair-addition and -removal strengths $B({\rm Pad/rm}0)$
of the ground-state transition, the excitation energy $E$ and the
pair-addition strength $B({\rm Pad}0)$ of the pair-addition vibrational
mode, and the excitation energy $E$ and the
pair-removal strength $B({\rm Pad}0)$ of the pair-removal vibrational mode
in Sn isotopes. The unit of the excitation energy is MeV. The ground-state
transitions in the closed shell nuclei $^{100}$Sn and $^{132}$Sn are
evaluated in the QRPA, separately for the pair-addition and pair-removal
transitions.
}
\end{table}


\subsection{Comparison with  pair rotational ground-state transition}

\begin{figure*}[t]
\includegraphics[scale=0.3,angle=-90]{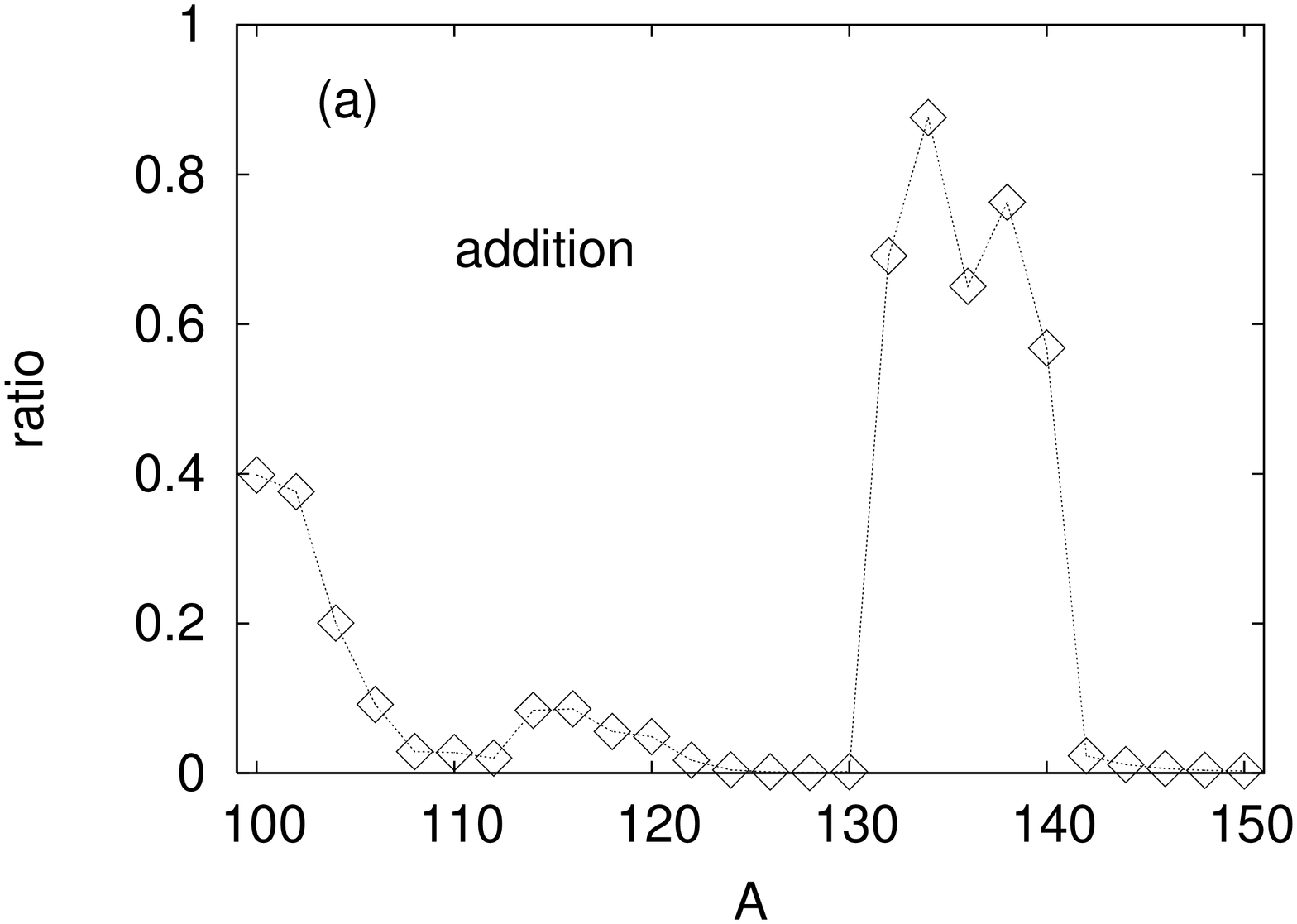}
\includegraphics[scale=0.3,angle=-90]{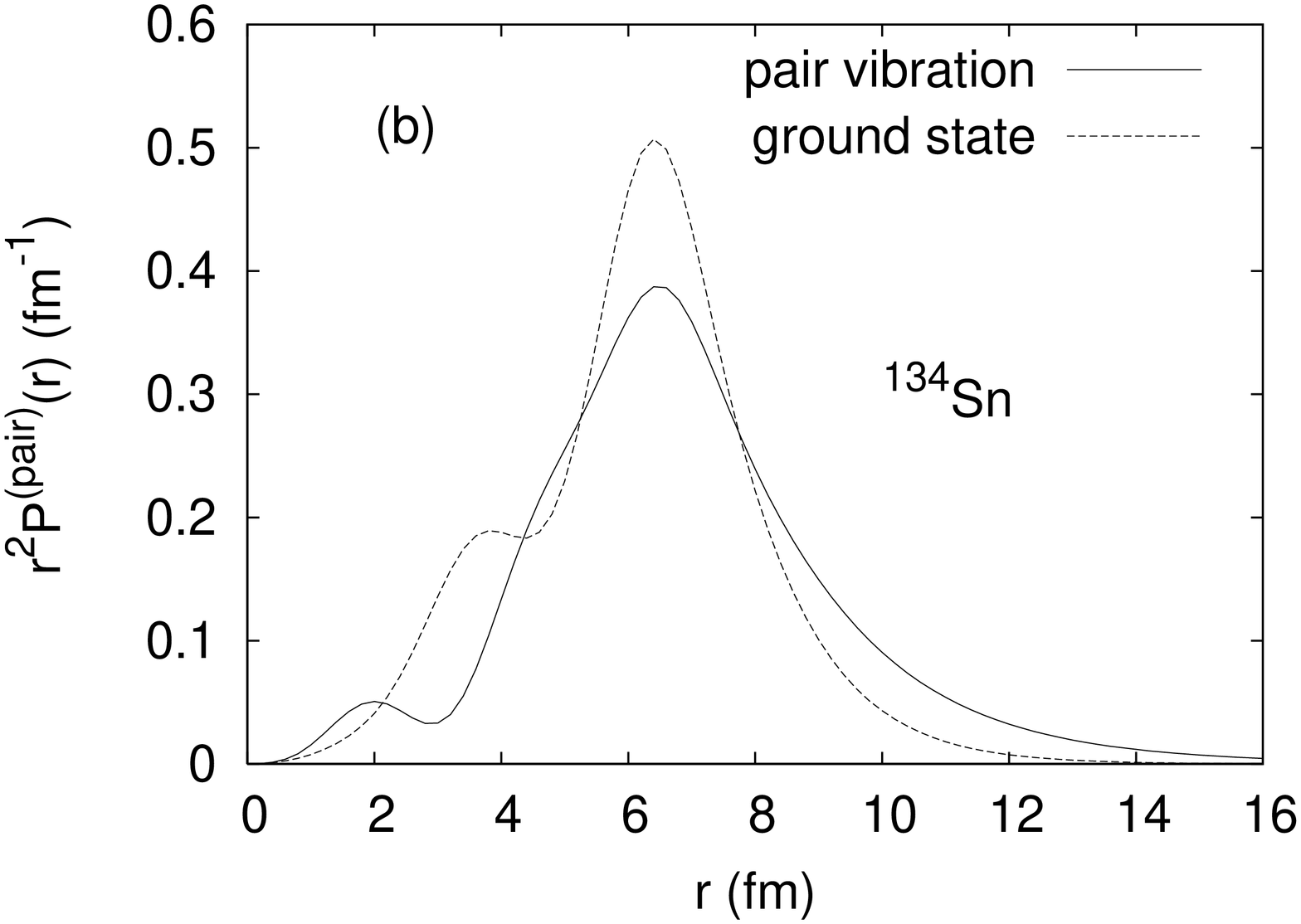}
\includegraphics[scale=0.3,angle=-90]{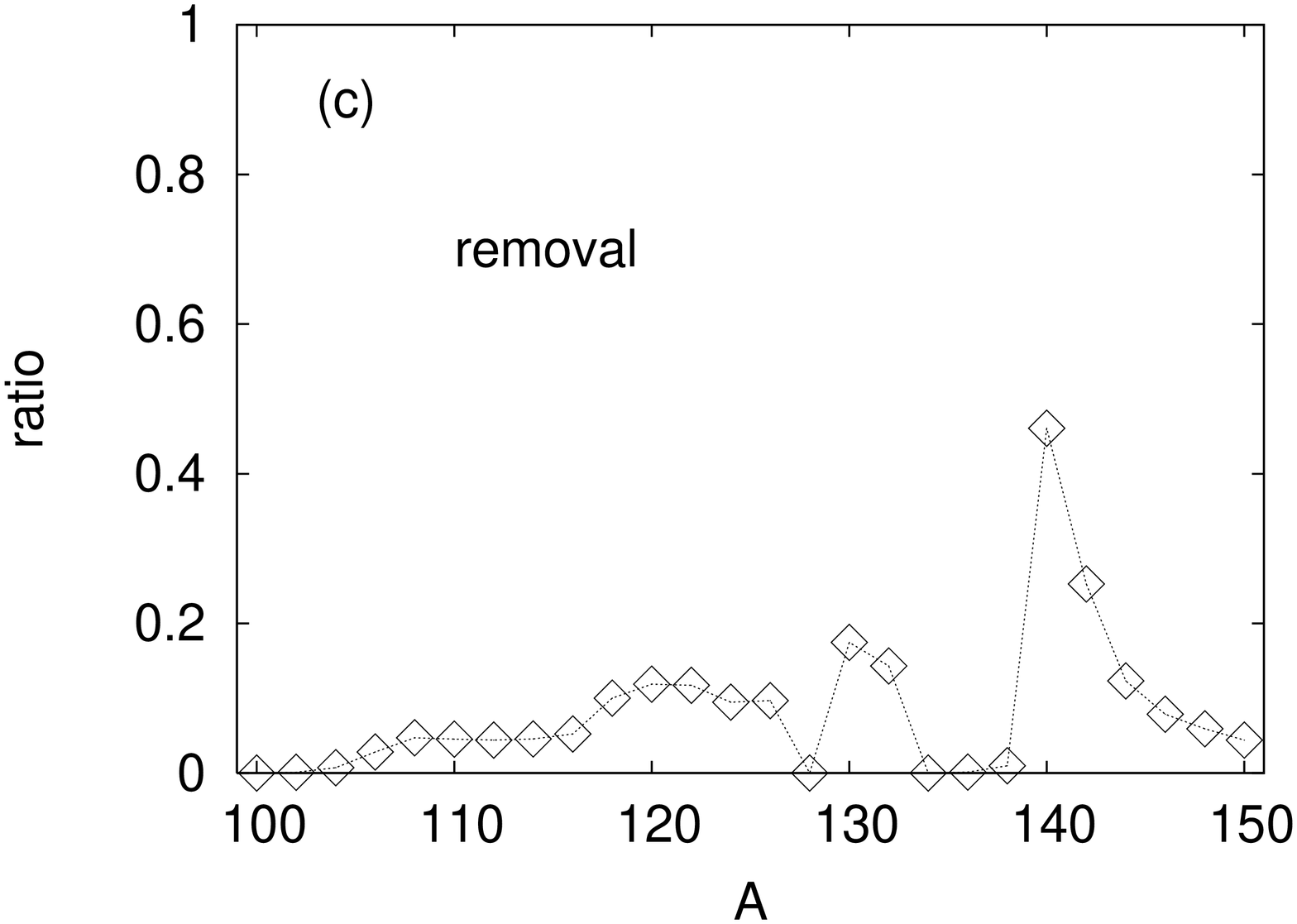}
\caption{\label{fig10}
(a) The ratio of the pair-addition strength $B({\rm Pad}0; {\rm gs} \rightarrow {\rm pv} )$ 
of the low-lying
 pair-addition
vibrational mode and the strength $B({\rm Pad}0; {\rm gs}\rightarrow {\rm gs})$ of 
the ground-state transfer.
(b) The pair-addition transition density $r^{2}P^{({\rm ad})}_{i L=0}(r)$ 
associated with the anomalous pair vibrational mode 
( solid curve)  and that for the ground-state transfer 
( dashed curve) in $^{134}$Sn.
(c) The ratio of the pair-removal strength 
$B({\rm Prm}0; {\rm gs} \rightarrow {\rm pv})$ of the low-lying
pair-removal mode and the ground-state transfer strength
$B({\rm Pad}0;{\rm gs} \rightarrow {\rm gs})$.
}
\end{figure*}

The anomalous feature of the pair vibrational modes in $A=132-140$ is clarified
further by comparing with the pair transfer populating the ground state,
i.e., the pairing rotation discussed in Section~\ref{sec:gs}.

Figure \ref{fig10}(a) show the ratio 
$r=B({\rm Pad}0;{\rm gs}\rightarrow {\rm pv})/B({\rm Pad}0;{\rm gs}\rightarrow {\rm gs})$ 
of the pair-addition strength 
associated with the low-lying pair vibrational mode and that with the pair
rotational ground-state transfer. 
The strength ratio amounts to 60-90 \% in the $A=132-140$ isotopes. 
It is
compared with isotopes 
in the middle of the $N=50-82$ shell, where the ratio is far below
10 \% (except in a few isotopes at the begging of the shell). 
The small ratio in the isotopes $A\sim 110-130$ is in qualitative agreement with the well established 
observation that two-neutron transfer cross sections populating low-lying $0^+$ states 
is generally weak,  
for instance, less than 10\% in stable Sn isotopes\cite{Broglia73,Brink-Broglia}. 
 A known exception
is the case  where an excited $0^+$ state having the character of the shape coexistence
emerges as a consequence of sudden shape changes with the 
neutron number\cite{Broglia73,Wood92}. 
The large strength ratio of 60-90 \% in the $A=132-140$ isotopes 
is comparable with the cases of the shape transition/
shape coexistence.

In Fig.~\ref{fig10}(b), we compare the pair-addition transition densities
$P^{({\rm ad})}_{i L=0}(r)$ for the ground-state transitions and for the
pair-vibrational transitions populating excited $0^+$ states. 
The maximal values of the amplitudes are
comparable, but 
 the transition density of the anomalous pair vibration mode 
extends more than that associated with the pair rotational 
ground-state transition. 
It reflects different microscopic structures
of the pairing rotation and of the anomalous pairing vibration. As we discussed
in connection
with Fig.~\ref{fig2}, the largest component of the pairing rotation may be 
 $[2f_{7/2}]^2$, while the counter part of the anomalous pairing vibration may be
 $[3p_{3/2}]^2$ and $[3p_{2/1}]^2$. The $3p$ orbits have much smaller
 binding energy and thus longer tail in the wave functions than those of
 $2f_{7/2}$.  The difference in the spatial extension
of the transition densities may be explained in this context.

The ratio $r'=B({\rm Prm}0;{\rm gs}\rightarrow {\rm pv})/B({\rm Prm}0;{\rm gs}\rightarrow {\rm gs})$ 
of the
strengths of the pair-removal vibrational mode and of the pairing rotation
is shown in Fig.~\ref{fig10}(c). 
The ratio in $^{140-144}$Sn is $30-40\%$, but it is relatively 
small if compared to the ratio $60-90 \%$ of the pair addition strengths in $^{132-140}$Sn,
and it decreases to small values in $A>146$.
Note that in the isotopes beyond $A=140$ the strengths of the  pair rotational ground-state 
transfer is 
more drastically increased than the strengths of the pair-removal 
vibrational mode.

\section{Sensitivity to density-dependent pairing}

\begin{figure*}[t]
\includegraphics[scale=0.3,angle=-90]{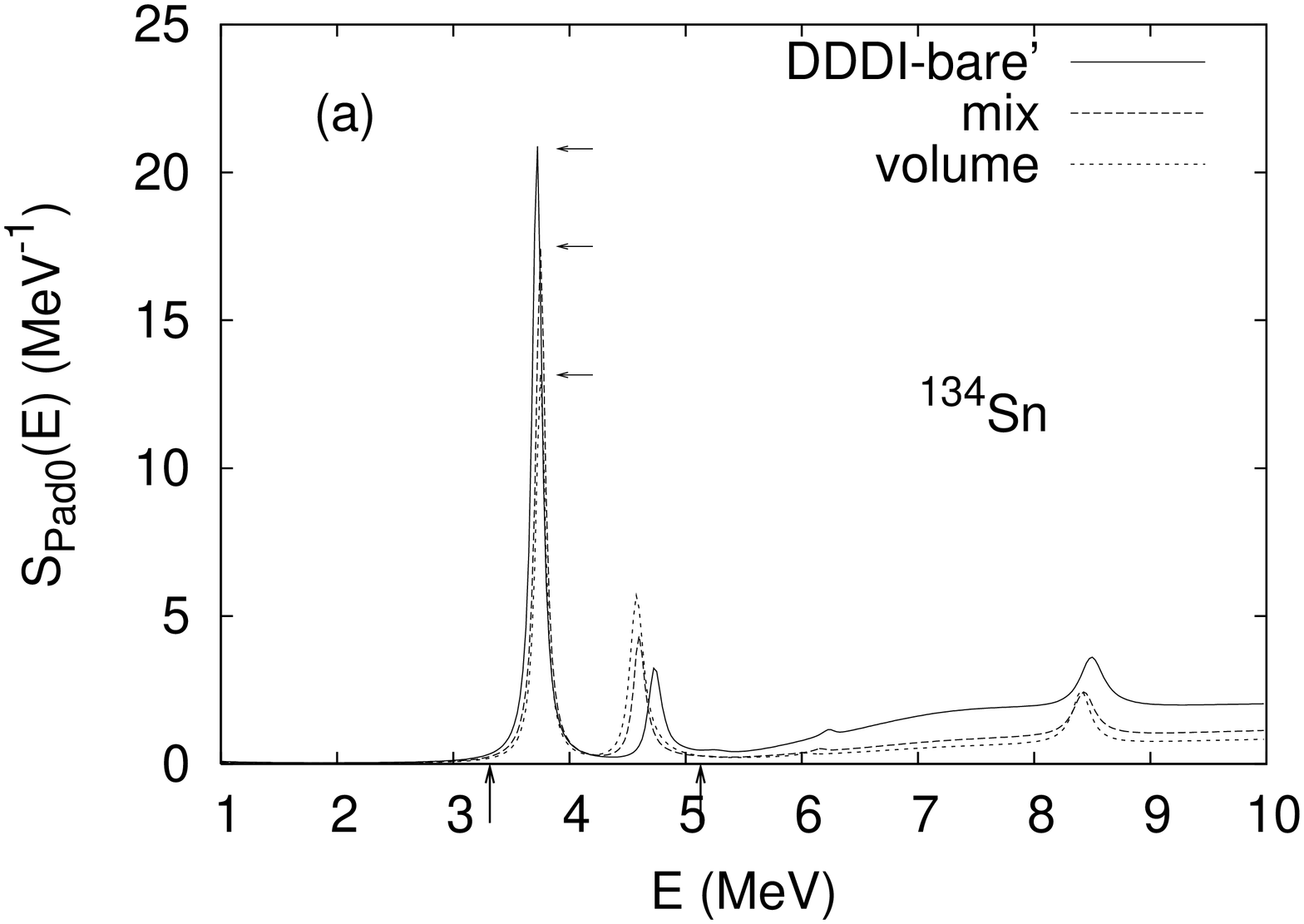}
\includegraphics[scale=0.3,angle=-90]{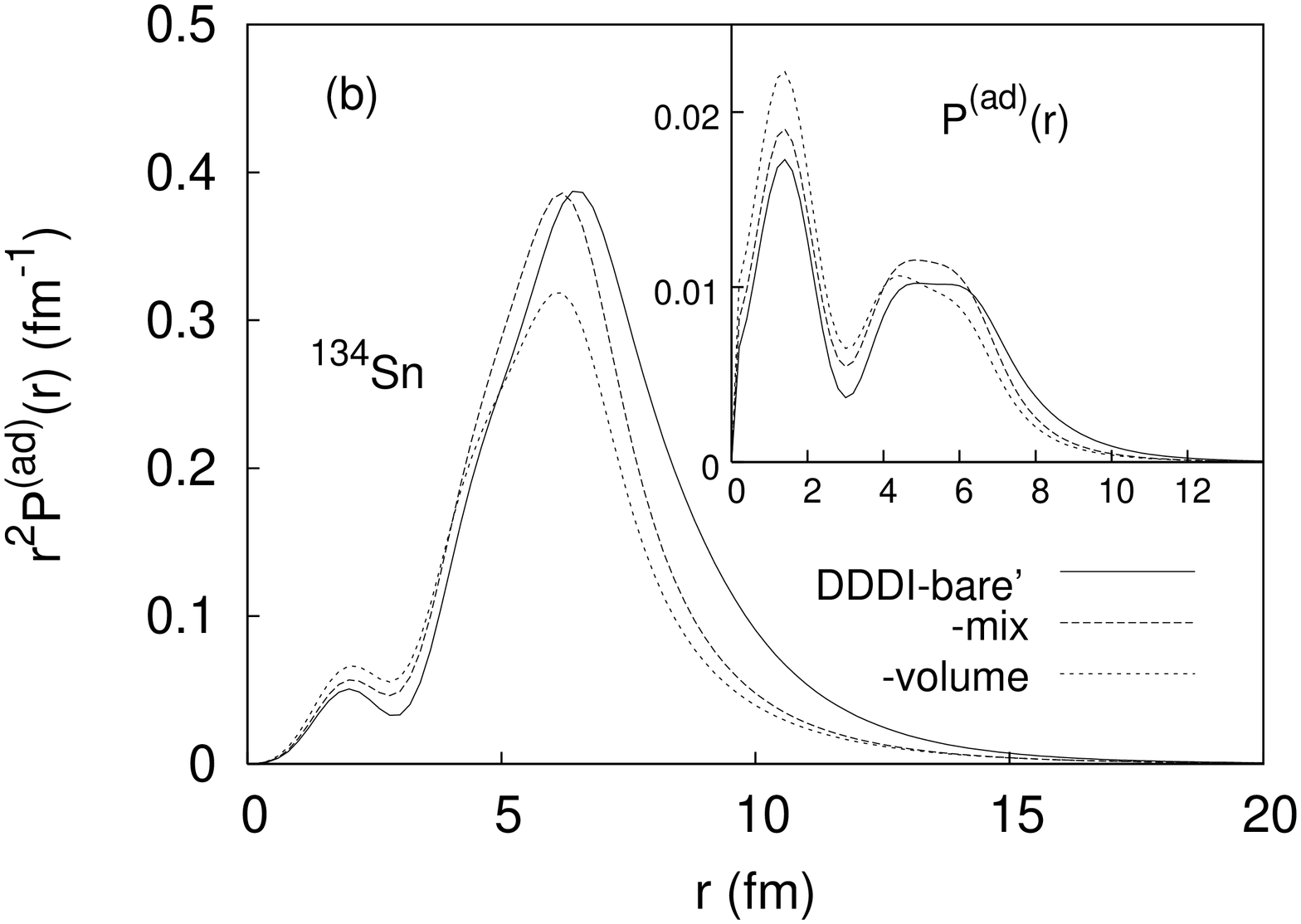}
\caption{\label{fig11}
(a) The pair-addition strength function $S_{{\rm Pad}0}(E)$ for neutrons in $^{134}$Sn, calculated with 
the three effective pairing interactions; the DDDI-bare', the mix, and the volume types.
The arrows indicate the peak heights.
(b) The transition density $r^{2}P^{({\rm ad})}_{i L=0}(r)$ of the low-lying pair-addition vibrational
mode in $^{134}$Sn, calculated with 
the three effective pairing interactions; the DDDI-bare', mix, and volume types.
In the inset we show the transition density $P^{({\rm ad})}_{i L=0}(r)$ without the
volume element.
}
\end{figure*}

It is interesting to examine the sensitivity of the pair transfers 
to different effective pairing interactions\cite{Khan09,Matsuo10}.  
For this purpose we have performed the HFB+QRPA calculations with
the three different versions
of the DDDI described in Section \ref{sec:model}. Note that 
the three interactions are designed to give approximately the same values of the
average neutron pair gap $\Delta_{uv}$
(cf. Fig.~\ref{fig1}(b)). Comparing the three interactions, we 
can examine effects of  different density-dependences
of the DDDI while keeping approximately the same average pair gap.

The influence of the density dependence on  the pair-addition and -removal strengths
$B({\rm Pad/rm}0)$ of the ground-state transition is shown in Fig.~\ref{fig1}(a). 
Although the influence is not significant for 
$A=100-132$, we see some sensitivity (but not very strong one) to the density dependence for neutron-rich
isotopes with $A>132$.

Concerning the pair vibrational modes, in contrast, 
the anomalous pairing vibration in  $A=132-140$ has significant
sensitivity to the density-dependence of the effective pairing interactions. 
As an example,
we show in Fig.~\ref{fig11}(a) the pair-addition strength function  $S_{{\rm Pad}0}(E)$
in $^{134}$Sn calculated with
the three different DDDI's. It is noticed  that
the basic structure of the strength function  
is the same, and
 the anomalous pair vibrational mode commonly exists at around $E\approx 3.5$ MeV. 
More importantly it is seen that the height of the pair vibrational peak depends
strongly on the different pairing interaction by about a factor of 1.5. 
The difference is  larger than  that found in the
ground-state transition. The same effect is
seen commonly at 
$A=132-138$ as shown in Fig.~\ref{fig7}(b). 
On the other hand, the influence of the density-dependence on the
pair-addition vibrational in the isotopes $A<132$ mode is small. 

We can understand the influence of the density-dependence on the anomalous
pair vibrational mode in terms of the  transition density
$P^{({\rm ad})}_{i L=0}(r)$, whose dependence on the pair interactions is shown in
Fig.~\ref{fig11}(b). It is seen in this figure that
the amplitude  $P^{({\rm ad})}_{i L=0}(r)$
in the external region $r \gesim R_{{\rm rms}}$ outside the surface
shows the same trend as in the pair-addition strength $B({\rm Pad}0)$;
the DDDI-bare' $>$ mix  $>$ volume pairing interactions.
We can then relate this trend to the features of density-dependent 
 interaction strength $V_{n}(r)$, which is, 
in the external region, the largest for the DDDI-bare'  and the smallest for the volume pairing,
reflecting the density-dependent interaction strength at low densities.
  ($v_0=-458,-292,-195$ MeV fm$^{3}$ and see also Fig.1 of Ref.\cite{Matsuo10} ). 
Since the anomalous pair vibrational mode at $A=132-140$ 
has the transition density extending far outside the surface,
the large difference in the interaction strength $V_{n}(r)$ in the exterior
is effective to this mode. 
However, 
for the usual pair vibrational modes whose
transition density does not reach very far, the sensitivity becomes small.

Finally we remark on related works\cite{Khan09,Matsuo10} discussing the
sensitivity to the density-dependence of the effective pairing interaction. 
In Ref.~\cite{Khan09},  the pair-transfer  strengths and the transition density 
of the pair vibrational modes are analyzed for  $^{124}$Sn and $^{136}$Sn by using the
Skyrme-HFB + QRPA model, which is similar to the model adopted here. In the case of
$^{136}$Sn, the low-lying pair vibrational mode around $E \sim 3$ MeV emerges in two of their
calculations (the $\eta=0.65$ and 0.35 cases of Ref.\cite{Khan09}), in qualitative 
agreement with our results. 
The shape of the pair-addition transition density $P^{({\rm ad})}_{i L=0}(r)$ 
is also similar as seen from the comparison of the inset of
Fig.~\ref{fig11}(b) of the present paper and Fig.8 of Ref.\cite{Khan09}. On the other hand, 
in another
calculation adopting the DDDI with the strongest density dependence (the $\eta=1$ case 
in Ref.\cite{Khan09}),
the low-lying pair vibrational mode 
is fragmented into three peaks  (cf. Fig.7 of  Ref.\cite{Khan09}), suggesting a complex sensitivity
to the pairing interactions.
We note that 
the interaction strength $v_0=-670$ MeVfm$^{3}$ chosen in the $\eta=1$ case is too large since
it leads to 
the positive scattering length, implying unphysical existence of a bound state for the $S=0$ neutron pair.
In the present investigation, on the contrary, we constrain the upper limit of  $|v_0|$ as 
$|v_0| \leq 458$ MeV fm$^{3}$ 
 by the scattering length $a=-18.5$ MeV
of the $^{1}S$ channel of the bare nuclear force. 
For our pairing interactions determined under this constraint,
 the anomalous pair vibrational mode stably
emerges  as a single-peak around $E \sim 3-4$  MeV  in $^{132-140}$Sn.

It is discussed in Ref.\cite{Matsuo10} that the quadrupole pair-addition transfer populating the first $2^+$ states
in $^{134}$Sn and heavier isotopes is sensitive to the density-dependence of the pairing interaction. In that case
the ratio of the pair-addition strengths  
between the volume pairing and the DDDI-bare'  amounts to approximately a factor of two, which is
slightly larger than the sensitivity (a factor of $\sim 1.5$) of the pair vibrational mode in $^{132-140}$Sn. 
The origin of the different sensitivity is not clear at present, but it may be related to the fact that the monopole
pair transition density has larger amplitude also in the internal region of the nucleus
than  that of the quadrupole mode. 

\section{Conclusions}

We have investigated microscopically monopole two-neutron transfer modes
 in heavy-mass 
superfluid nuclei  using the  Skyrme-Hartree-Fock-Bogoliubov mean-field
 model  and the continuum quasiparticle random phase approximation. 
Emphases have been put 
 on  the pair vibrational modes populating  low-lying collective $0^+$ states and 
on the pair rotational 
transitions  connecting the ground states of the neighboring $\Delta N=\pm 2$ isotopes.
We have performed systematic numerical analysis  
for the even-even Sn isotopes ranging from $A=100$ to $A=150$, and  found
the following new  features of the two-neutron transfer modes emerging
in neutron-rich
Sn isotopes beyond the $N=82$ shell closure ($A>132$).

The calculation predicts  a novel type of pair vibrational mode in the isotopes $^{132-140}$Sn.
It is a monopole vibrational mode characterized by 
intense transition from the ground state to a low-lying $0^+$ state in the neighboring $N+2$ isotope 
via the two-neutron addition transfer. 
The corresponding $0^+$ states 
are predicted to emerge near the threshold energy of the one-neutron separation,
and to form a narrow resonance in case when it is located above the separation energy.  
An  anomalous feature of this pair vibrational mode and a marked 
difference from the pairing vibration in nuclei near the stability line
is that the pair transition
density exhibits a large amplitude in the
region outer than the surface, and also a long tail extending
up to $r\sim 15$ fm. We expect that the difference
may manifest itself  in
reaction observables such as the cross section of the
$(t,p)$ reaction although quantitative analysis remains for
future investigations.
As a microscopic origin of the anomalous pair vibrational mode, the
weakly bound or resonant $3p_{3/2}$ and $3p_{1/2}$ orbits 
are suggested to play key roles. 
 
In very neutron-rich isotopes with $A>140$ ($N>90$), enhancement of
the two-neutron transfer strength is predicted for the 
 transitions between the ground states. This is a natural 
consequence of the spatially extended pair field originating from
the weak binding of neutrons. Particularly, occupation of
the
 weakly bound  $3p_{3/2}$ and $3p_{1/2}$ orbits gives the sudden increase
 in the strength for $A>140$.  In this perspective, the anomalous pair vibrational 
mode in $^{132-140}$Sn can be linked to the strong ground-state transfer
in $A>140$, and it 
can be regarded as a precursor of the enhanced pair transfer which 
emerges in weakly bound neutron-rich nuclei close to the drip-line.

\begin{acknowledgments}
This work was supported by
the Grant-in-Aid for Scientific Research 
(Nos. 20540259, 21105507, 21340073 and 23540294) from the Japan
 Society for the Promotion of Science,   and also by
the JSPS Core-to-Core Program, International
Research Network for Exotic Femto Systems(EFES).

\end{acknowledgments}

\end{document}